\pdfoutput=1
\documentclass[iop]{emulateapj}
\usepackage{graphicx}
\usepackage{epsfig}
\usepackage{epstopdf}
\usepackage{amssymb}
\usepackage{multirow}
\usepackage{ulem}
\newcommand{\Se}{Section}
\def\kms{$\textrm{km~s$^{-1}$}$}

\def\H2{H$_{2}$}

\def\roH2{$\rho_{\textrm{H}_2}$}
     
\def\MH2{M$_{\textrm{H}_2}$}

\def\Hb{H$\beta$ \,} 
\def\i{\,{\small I}}
\def\ii{\,{\small II}} 
\def\iii{\,{\small III}}

\shorttitle{Isolated S0 galaxies} \shortauthors{Katkov et al.}

\begin{document}

\title{Kinematics and stellar population in isolated
lenticular galaxies\footnotemark[1]} 
\footnotetext[1]{Based on observations made with
the Southern African Large Telescope (SALT), programs
\mbox{2011-3-RSA\_OTH-001}, \mbox{2012-1-RSA\_OTH-002} and
\mbox{2012-2-RSA\_OTH-002}.}

\author{Ivan Yu. Katkov} 
\affil{Sternberg Astronomical Institute, M.V. Lomonosov Moscow State University, Universitetsky pr., 13, Moscow, 119991 Russia}
\email{katkov@sai.msu.ru} 

\author{Alexei Yu. Kniazev}
\affil{South African Astronomical Observatory, PO Box 9, 7935 Observatory, Cape Town, South Africa}
\affil{Southern African Large Telescope Foundation, PO Box 9, 7935 Observatory, Cape Town, South Africa}
\affil{Sternberg Astronomical Institute, M.V. Lomonosov Moscow State University, Universitetsky pr., 13, Moscow, 119991 Russia}
\email{akniazev@saao.ac.za}

\author{Olga K. Sil'chenko} 
\affil{Sternberg Astronomical Institute, M.V. Lomonosov Moscow State University, Universitetsky pr., 13, Moscow, 119991 Russia}
\affil{Isaac Newton Institute, Chile, Moscow Branch}
\email{olga@sai.msu.su} 

%\and 

\begin{abstract}

By combining new long-slit spectral data obtained with the
Southern African Large Telescope (SALT) for 9 galaxies with previously
published our observations for additional 12 galaxies we study the stellar and
gaseous kinematics as well as radially resolved stellar population properties
and ionized-gas metallicity and excitation for a sample of isolated lenticular
galaxies. We have found that there is no particular time frame of formation
for the isolated lenticular galaxies: the mean stellar ages of the bulges and
disks are distributed between 1 and $>13$~Gyr, and the bulge and the disk in
every galaxy formed synchronously demonstrating similar stellar ages and
magnesium-to-iron ratios. Extended ionized-gas disks are found in the majority
of the isolated lenticular galaxies, in 72\%$\pm 11$\%. The half of all
extended gaseous disks demonstrate visible counterrotation with respect to
their stellar counterparts. We argue that just such fraction of projected
counterrotation is expected if all the gas in isolated lenticular galaxies is
accreted from outside, under the assumption of isotropically distributed
external sources. A very narrow range of the gas oxygen abundances found by us
for the outer ionized-gas disks excited by young stars, [O/H] from 0.0 to $+0.2$~dex, 
gives evidence for the satellite merging as the most probable source of this
accretion. At last we formulate a hypothesis that morphological type of a
field disk galaxy is completely determined by the outer cold-gas accretion
regime.

\end{abstract}

\keywords{galaxies: elliptical and lenticular - galaxies: evolution -
galaxies: formation - galaxies: kinematics and dynamics - galaxies:
structure.}

\section{Introduction} 
Understanding processes of galaxy formation and evolution is the greatest
challenge  for modern extragalactic astrophysics. A huge variety of physical
processes are involved into galaxy shaping, and it is a key problem to select
the dominant agents driving evolution of galaxies of different morphological
types.

Lenticular galaxies were introduced by Edwin \citet{hubble_1936} when
proposing his famous morphological scheme, `Hubble's fork', as a hypothetical
intermediate type between ellipticals (to the left) and spirals (to the
right): the so called S0s placed by him in the centre of the `Hubble's fork'
were to possess large-scale stellar disks as spiral galaxies, but lacked patchy
patterns of HII-regions and spiral arms unlike them. Their smooth reddish
appearance implied their resemblance to elliptical galaxies as concerning the
stellar populations ages. It was also long been suggested that the
intermediate position of the S0s between pure spheroids (elliptical galaxies)
and disk-dominated late-type spiral galaxies obliged them to have very large
bulges. However direct photometric decompositions of the digital images of a
representative sample of lenticular galaxies has proved that the bulges of any
size can be met in S0s, from very large to tiny ones \citep{laurikainen_2010}.
So the early idea of Sidney \citet{van_den_bergh_1976} that within the
Hubble's morphological scheme the S0 galaxies constitute in fact a
morphological sequence parallel to the sequence of spiral galaxies, matching
their bulge-to-disk ratios at every morphological subtype, becomes now more
and more acceptable \citep{kormendy_bender_2012, cappellari_atlas7_2011}. At
first glance, this re-forming of the Hubble morphological sequence strengthens
the general opinion that S0s may be formed by quenching star formation in the
disks of spiral galaxies -- this transformation step must be easier to do when
the bulge-to-disk ratios are the same in the progenitor and the descendant.
But at this point we would like to note that if the bulges of S0s and spirals
are indeed similar, the possibility of obtaining a {\it spiral} galaxy from a
{\it S0 progenitor} arises while this direction of evolution was excluded when
the bulges of S0s were supposed to be systematically larger than the bulges of
spirals.

A great variety of physical processes that can in principle quench star
formation in a disk of a spiral galaxy, to transform it into a lenticular one,
are currently discussed: a very incomplete list includes direct collisions
\citep{spitzer_1951, icke_1985}, tides from a cluster/group dark halo
potential \citep{byrd_1990}, `harrasment' -- high-speed encounters between
galaxies in dense environments \citep{moore_1996}, ram pressure by the hot
intergalactic medium \citep{gunn_gott_1972, quilis_2000}, starvation of star
formation because of removing external gas reservoir \citep{larson_1980,
shaya_tully_1984}.  All these processes are inevitably related to dense
environments: it is suggested that only clusters and rich groups of galaxies,
having massive dark haloes enclosing many individual galaxies, can provide
necessary density of intergalactic medium for effective ram pressure and
galaxy tight packing for effective gravitational tides. On one hand, indeed,
S0 galaxies are known to be the dominant galaxy population of nearby clusters,
their fraction in clusters reaching 60\%\ \citep{dressler_1980}. On the other
hand, there is an even larger number of S0 galaxies in the field: the galaxy
content of the nearby Universe includes 15\%\ of lenticulars
\citep{naim_1995}. Some quite isolated lenticular galaxies even exist
\citep{sulentic_2006}. What are the mechanisms of their formation? Can be they
quite different from those acting in dense environments? This question has not
even been considered.

Despite the obvious scarcity of possible galaxy transformation mechanisms
beyond the clusters and rich groups, it would be erroneous to think that an
isolated galaxy evolves as a `closed box'.  Recently we have studied an
isolated early-type spiral galaxy NGC~7217. By analysing a full complexity of
its properties including disk structure, dynamical state, inner gas polar
disk, and stellar population characteristics along the radius, we have shown
that its present structure requires at least two satellite infalls (minor
merging) for the last 5~Gyr \citep{sil_2011_n7217}. A noticeable gas presence
in S0s is not rare, and in particular, off-cluster environments have appeared
to favor an external (accretion) origin of this gas \citep{davis_2011}.
Moreover, we have shown that in extremely sparse environments, namely, in the
quite isolated S0s the warm extended gas is {\it always} accreted from outside
\citep{ilg_gas}. So external gas acquisition, related to smooth cold-gas
accretion and/or to merging small late-type gas-rich satellites, together with
the inner cold-disk instabilities, remain the only possible drivers of
isolated lenticular galaxy formation and evolution. By taking this idea in
mind, we have undertaken a study of the kinematical and stellar population
properties of {\it isolated} lenticular galaxies, by hoping to single out the
evolutionary paths related just to the gas/satellite accretion regime. Here we
must also note that the gas accretion and/or minor merging allowed for the
field disk galaxies are able not only to quench star formation in a large-scale 
disk, but to feed and provoke it in the disks where it has not proceeded
before (e.g. \citet{birnboim07, sancisi_rev}).

To achieve this goal, we have compiled a list of strictly isolated nearby
($v_r<4000$ \kms) lenticular galaxies and have undertaken deep long-slit
spectroscopy of a small representative sample of them to study the kinematics
of the stars and of the gas and the ages and chemical compositions of the
stellar populations as well as the ionization mechanisms and metallicity of
the warm-gas component. With these results in hands, we hope to restore
formation and evolutionary histories of the isolated lenticular galaxies. In
this paper we show and analyze the data on the southern part of the sample:
the galaxies having been observed at the Southern Africa Large Telescope
(SALT) are presented. The paper is organized as follows: \Se~\ref{txt:Sample}
describes the sample, \Se~\ref{txt:Obs_and_Red} gives the description of the
observations and data reduction, in \Se~\ref{txt:results} we present our
results and discuss them in \Se~\ref{txt:discus}, the conclusions drawn from
this study are summarized in \Se~\ref{txt:summ}.

\section{Sample selection}
\label{txt:Sample}

Our approach to compile a sample of strictly isolated lenticular galaxies is
based on a set of methods developped in the Laboratory of Extragalactic
Astrophysics and Cosmology of the Special Astrophysical Observatory of the
Russian Academy of Sciences by Igor Karachentsev, Dmitry Makarov, and their
coauthors. They proposed a new group-finding algorithm which was intended to
be applied to their Nearby Galaxy Catalog \citep{Karachentsev_neigb_gal_2004,
Karachentsev_neigb_gal_update_2013}.  By extending their study of the local
large-scale structures, they have also used their algorithms to identify
galaxy pairs \citep{karachentsev_pairs_2008}, triplets
\citep{makarov_triplets_2009}, groups \citep{makarov_groups_2011}, and
isolated galaxies \citep{karachentsev_isol_2011} up to Hubble velocities of
$v_r\le 4000$ \kms. The updated HyperLEDA and NED databases extended by
measurements coming from the surveys SDSS, 6dF, HIPASS, and ALFALFA, provided
line-of-sight systemic velocities, apparent magnitudes, and morphological
types of the galaxies under consideration. The profit of their group-finding
approach is that the individual properties of galaxies, in particular an
integrated luminosity in the $K$-band as a stellar mass proxy, are taken into
account. They assumed that velocity difference and visible separation of
galaxies belonging to a physical pair must both satisfy the condition of
negative total energy, and the pair components must be enclosed within the
sphere of `zero-velocity' that means that the pair components are separated
from the Hubble expansion flow.  This algorithm for galaxy grouping is
iterative: galaxy-galaxy physical pairs are identified during the first
iteration, and after that at the subsequent iterations the galaxy pairs are
tied into groups through the common members. The isolation index ($II$)
characterizing isolation degree of any galaxy within the sphere of $v_r\le 4000$
\kms\ is a by-product of all the galaxy grouping procedures. The $II$ value of
an unbound galaxy pair is larger than one and indicates a factor by which the
mass of one of the components should be increased in order to create a
gravitationally bound pair. Correspondingly, the $II$ values of the galaxies
belonging to multiple systems are less than one.

Dmitry Makarov has kindly provided us with the information about the isolation
indices for all galaxies of the Local Supercluster and its surroundings. To
define our sample of isolated lenticular galaxies, we have selected early
morphological types, $-3 \le T \le 0$, with the isolation indices $II > 2.5$.
Also we have taken some galaxies having faint companions with $1< II < 2.5$
but with the $K$-magnitude difference of 3 mag and larger having in mind that
the low-mass satellites, with the mass of ten percent and less relative to
their host, cannot gravitationally affect the evolution of their hosts (unless
they merge). The whole sample of the isolated lenticular galaxies, both of the
northern and southern skies, lists 281 objects. We have started spectral
observations of a representative part of this sample.  Firstly, we have
carried out spectroscopic observations of 12 northern targets from the sample
of isolated S0 galaxies at the 6-m Russian telescope by using universal
SCORPIO spectrograph; the results are published in \citet{ilg_gas,
katkov_ilg_stpop}. In this paper we present results of the long-slit
spectroscopic study of 9 targets of the southern hemisphere undertaken at the
Southern African Large Telescope (SALT). As a Discussion, some summary of the
results for the unified sample of northern and southern isolated S0s is also
given at the end of the paper. 

%\placetable{table_env}

\section{Long-slit spectroscopy}
\label{txt:Obs_and_Red}
\subsection{Observations and data reduction}

The observations were performed with the Robert Stobie Spectrograph
\citep[RSS;][]{Burgh03,Kobul03} at the Southern African Large Telescope (SALT)
\citep{Buck06,Dono06}. The long-slit spectroscopy mode of the RSS was used with
a 1.25 arcsec slit width for the most observations.  The total time of one
observational block with the SALT is limited by the track-time of about an hour
for our targets.  For this reason and because the SALT is a queue-scheduled
telescope, most of our galaxies were observed more than once and all
observations were done during different nights.  All observational details are
summarised in Table~\ref{tbl_logobs}.  The slit was oriented along the major axis
for every galaxy except NGC~7693.  The grating GR900 was used for our program to cover 
finally the spectral range of 3760$-$6860 \AA\ with a final reciprocal dispersion of
$\approx0.97$ \AA\ pixel$^{-1}$ and FWHM spectral resolution of 5.5 \AA.  The
seeing during observations was in the range 1.5$-$3.0 arcsec.  The RSS pixel
scale is 0\farcs1267, and the effective field of view is 8\arcmin\ in diameter.
We utilised a binning factor of 2 or 4 to give final spatial sampling of
0\farcs258 pixel$^{-1}$ and 0\farcs507 pixel$^{-1}$ respectively.  Spectrum of
an Ar comparison arc was obtained to calibrate the wavelength scale after each
observation as well as spectral flats were observed regularly to correct for
the pixel-to-pixel variations.  Spectrophotometric standard stars were observed
during twilights, after observations of objects, for the relative flux
calibration.

% Table 2 %%%%%%%%%%%%%%%%%%%%%%%%%%%%%%%%%%%%%%%%%%%%%%%%%%%%%%%%%%%%%%%%%%%%%%%%%%%%%%%%%%%%

\begin{deluxetable*}{cclccccc}
\tablecolumns{8}
\tablewidth{0pc}

\tablecaption{Long-slit spectroscopy of studied galaxies.\label{tbl_logobs}}

\tablehead{
\colhead{Galaxy}       & \colhead{Date}    &  \colhead{Exp.}      & \colhead{Binning}    & \colhead{Slit}    & 
\colhead{PA(slit)} & \colhead{Seeing}          \\
             &           & \colhead{[sec]}      &            & \colhead{[arcsec]} &  \colhead{[deg]}   & \colhead{[FWHM, arcsec]}  
}         

\tabletypesize{\scriptsize}    

\startdata

\multirow{2}{*}{IC\,1608}  & 07.11.2012 & 620x3      & 2$\times$4 &  1.25   & 350 & 2.0 \\
                           & 04.01.2013 &  620x3      & 2$\times$4 &  1.25   & 350 & 2.0 \\ \hline
%
%07.01.2013 & IC\,3152  & 600x3      & 2$\times$4 &  1.25 0 & 223 & 2.0 \\
%14.01.2013 & IC\,3152  & 600x3      & 2$\times$4 &  1.25 0 & 223 & 2.0 \\ \hline
%
IC\,4653                   & 11.05.2012 &  1200x3     & 2$\times$2 &  1.25   & 52 & 2.0 \\ \hline
\multirow{4}{*}{NGC\,1211} & 05.10.2011 & 820x3      & 2$\times$2 &  1.25   & 210      & 3.0 \\
                           & 22.11.2011 & 900x3      & 2$\times$2 &  1.25   & 210 & 3.0 \\
                           & 22.12.2011 & 1030x2,730 & 2$\times$2 &  1.25   & 210 & 3.0 \\
                           & 25.12.2011 & 1000x3     & 2$\times$2 &  1.25   & 210 & 3.0 \\ \hline
\multirow{4}{*}{NGC\,2917} & 17.12.2012 & 900x2      & 2$\times$4 &  1.25   & 169 & 3.0 \\
                           & 06.01.2013 & 900x3      & 2$\times$4 &  1.25   & 169 & 3.0 \\
                           & 15.01.2013 & 900x3      & 2$\times$4 &  1.25   & 169 & 3.0 \\
                           & 15.02.2013 & 900x2,700  & 2$\times$4 &  1.25   & 169 & 3.0 \\ 
\multirow{3}{*}{NGC\,3375} & 17.02.2012 & 850x2,472  & 2$\times$2 &  1.25   & 130 & 2.0 \\
                           & 23.02.2012 & 800x3      & 2$\times$2 &  1.25   & 130 & 2.0 \\
                           & 28.02.2012 & 800x3      & 2$\times$2 &  1.25   & 130 & 2.0 \\ \hline
\multirow{2}{*}{NGC\,4240} & 14.01.2013 & 600x3      & 2$\times$4 &  1.25   & 283 & 2.0 \\
                           & 19.03.2013 & 600x3      & 2$\times$4 &  1.25   & 283 & 2.0 \\ \hline

NGC\,6010                  & 05.04.2013 & 750x3      & 2$\times$4 &  1.00   & 105 & 2.0 \\ \hline
\multirow{3}{*}{NGC\,7693} & 10.07.2012 & 650x3      & 2$\times$2 &  1.25   & 210 & 2.0 \\
                           & 04.09.2012 & 650x3      & 2$\times$2 &  1.25   & 30 & 2.0 \\
                           & 22.09.2012 & 650x3      & 2$\times$2 &  1.25   & 30 & 2.0 \\ \hline
\multirow{2}{*}{UGC\,9980} & 10.06.2012 & 700x3      & 2$\times$2 &  1.25   & 175 & 2.0 \\
                           & 10.07.2012 & 650        & 2$\times$2 &  1.25   & 175 & 2.0 \\ %\hline
\enddata
\end{deluxetable*}

Primary data reduction was done with the SALT science pipeline \citep{Cr2010}.
After that, the bias and gain corrected and mosaicked long-slit data were
reduced in the way described in \citet{Kn08}.  The accuracy of the spectral
linearisation was checked using the sky line [O\i]~$\lambda$5577; the RMS
scatter of its wavelength measured along the slit is 0.04~\AA.  The slit length
is approximately 8\arcmin, so sky spectra from the slit edges were used to
estimate the background during the galaxy exposures.

\subsection{The Lick index system at the SALT/RSS}

To derive stellar population properties from the integrated absorption-line
spectra of a stellar system, in particular of a galaxy or its part, one can use
the equivalent widths (EWs) of the stellar absorption spectral lines. Lick
indices \citep{Faber_1985, licksystem1, Worthey_Ottaviani_1997} is a uniform,
strictly established system of set line parameters measured in part as EWs of
strong absorption lines in the spectral range of 4000-6200~\AA. The system is
named after a 20-yr spectral survey of nearby galaxies and stars with the 3-m
Lick telescope using a photon-counting detector IDS at the Cassegrain
spectrograph. The line and continuum border definitions within the Lick system
are tied to the spectral resolution of the Lick spectrograph, $\approx$8~\AA\
but slightly varying with wavelength.  The aim was to include the spectral
lines fully into the integrated spectral ranges.  The necessity to apply the
Lick definitions of the absorption-line EW measurements to galactic spectra was
strengthened by the fact that many evolutionary synthesis models of simple
stellar populations, starting from the work of \citet{licksystem1}, used
calibrations of the {\it stellar} Lick indices on the stellar effective
temperatures and metallicities as their input data.  These calibration
relations were obtained from observations of more than 460 nearby stars with
exactly the same Lick spectral setup.

The Volume Phase Holographic (VPH) grating of 900 g/mm of the SALT/RSS has a
spectral resolution of about 5.5~\AA\, that differs from the standard Lick
resolution, $\approx 8$~\AA. We hence need to calibrate the instrumental
absorption-line indices obtained from the RSS spectra by integrating the
spectral fluxes in the bands prescribed by the Lick system to the standard Lick
system.  This was done by observing  a sample of Lick standard stars visible in
southern sky from the list of \citet{licksystem1}. In total, 10 giant and dwarf
stars with spectral types in-between F4 and K4 were observed with the VPH900
grating and the slit width of 1.25 arcsec. All observations of these bright
stars were done either within twilight time or during bright moon time with
poor seeing and cloudy conditions. For all obtained spectra we calculated the
instrumental-system Lick indices H$\beta$, Mgb, Fe5270, and Fe5335, by
integrating fluxes within the prescribed wavelength intervals for the lines, as
well as their blue and red continuum points, as recommended by
\citet{licksystem1}.

% Figure 1 %%%%%%%%%%%%%%%%%%%%%%%%%%%%%%%%%%%%%%%%%%%%%%%%%%%%%%%%%%%%%%%%%%%%%%%%%%%%%%%%%%%%%%%%%%%
\begin{figure}
%\resizebox{0.7\hsize}{!}{\includegraphics{katkov_fig1}}
\includegraphics[width=0.55\textwidth]{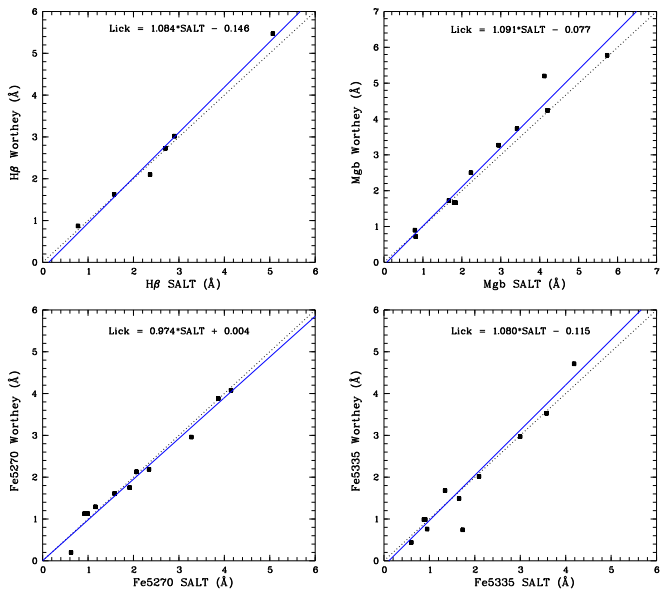}
\caption{The calibration of the instrumental Lick indices of the RSS/SALT
with grating VPH900 onto the standard Lick system. The straight solid lines are
the best-fit relations while the dashed lines are the equality relations.
}
\label{indsystem}
\end{figure}

The instrumental-system Lick indices were then compared to the tabular
values provided by \citet{licksystem1}. The linear dependencies between the
two sets of data were recovered, and the linear regressions were
calculated, and are also shown in Fig.~\ref{indsystem}:

$$
\mbox{H} \beta (\mbox{Lick}) = (1.084\pm 0.060) \times \mbox{H} \beta
(\mbox{\small RSS}) - (0.146\pm 0.158),
$$
\noindent
the rms scatter of the points around the straight line is 0.20~\AA.

$$
\mbox{Mgb} (\mbox{Lick}) = (1.091\pm 0.069) \times \mbox{Mgb}
(\mbox{\small RSS}) - (0.077\pm 0.189),
$$
\noindent
the rms scatter of the points around the straight line is 0.34~\AA.

$$
\mbox{Fe5270} (\mbox{Lick}) = (0.974\pm 0.053) \times \mbox{Fe5270}
(\mbox{\small RSS}) + (0.004\pm 0.113),
$$
\noindent
the rms scatter of the points around the straight line is 0.20~\AA.

$$
\mbox{Fe5335} (\mbox{Lick}) = (1.080\pm 0.059) \times \mbox{Fe5335}
(\mbox{\small RSS}) - (0.115\pm 0.116),
$$
\noindent
without HD\,10700, with the remaining 9 stars, the rms scatter
of the points around the straight line is 0.22~\AA.

Comparing the derived rms scatter of the individual stars around the
best-fit straight lines with the mean accuracy of the tabular Lick indices
mentioned by \citet{licksystem1}, namely 0.22~\AA\ for H$\beta$, 0.23~\AA\
for Mgb, 0.28~\AA\ for Fe5270, and 0.26~\AA\ for Fe5335, we conclude
that the scatter of the points in the plots of Fig.~\ref{indsystem}
is produced mostly by the errors of the tabular indices.

%{\bf Description of LSF determination based on Lick stars and reference frames!}

\subsection{Full spectral fitting}

Besides the Lick index measurements, we have also applied full spectral fitting
approach to our spectra; it is valuable when strong emission lines are present
in spectra, and the age-sensitive index H$\beta$ is strongly contaminated by
the Balmer emission.

In order to perform full spectral fitting of the synthetic spectra to the
observed data, we have used an \textsc{IDL}-based package \textsc{NBursts}
\citep{nbursts_a,nbursts_b}. This package implements a pixel-space fitting
algorithm, that involves the non-linear least-squares optimization using
Levenberg-Marquardt algorithm. The observed spectrum is approximated by a
stellar population model broadened by line-of-sight velocity distribution
(LOSVD); the parameters of the stellar population model, metallicity and age,
are determined during the same minimization loop as the internal kinematical
parameters -- line-of-sight velocity and stellar velocity dispersion. 

In our study, we use intermediate spectral resolution (R=10000) simple stellar
population (SSP) models generated by evolutionary synthesis code
\textsc{pegase.hr} \citep{pegasehr} in a wavelength range 3900-6800 \AA\ for
the \citet{SalpeterIMF} initial mass function based on ELODIE3.1 stellar
library \citep{elodie3.1}. The grid of synthetic SSP spectra was pre-convolved
with spectral line spread function (LSF) of the RSS spectrograph, which was
determined by fitting spectrum of one of Lick standard stars against the
R=10000 spectrum for the same star taking from ELODIE3.1 library. During the
main minimization loop the template spectrum is extracted from the grid of
models by interpolation to the current age $T$ and metallicity [Z/H].  Then
template is convolved with LOSVD, which is defined by Gauss-Hermite series of
orthogonal functions with parameters $v$, $\sigma$, $h_3$, $h_4$
\citep{gausshermite}. The model includes a multiplicative continuum aimed to
take into account flux calibration uncertainties both in observations and in
the models as well as possible dust attenuation of galactic spectrum.

In order to avoid systematic errors in the solution we masked narrow 15 \AA --
wide regions around ionized-gas emission lines and around traces of the
subtracted strong airglow lines. As shown by \citet{nbursts_a} and \citet{chil09_virgo},
excluding age-sensitive Balmer lines from the full spectral fit neither biases
age estimates nor significantly degrades the quality of the age determination.
To achieve the required signal-to-noise ratio of 20-30 per spatial bin, we
performed adaptive binning of the long-slit spectra along the slit. 

A number of similar approaches of the full spectral fitting techniques exist, for instance
\textsc{ppxf} by \citet{ppxf}, \textsc{starlight} by \citet{starlight},
\textsc{steckmap} by \citet{steckmap}, and other packages. The main difference between
the \textsc{NBursts} package and the current version of the popular \textsc{ppxf} code
as well as \textsc{starlight} and \textsc{steckmap} is that the \textsc{NBursts}
specifies template spectrum as a single SSP spectrum with age and metallicity
as free parameters. Other packages construct template spectrum as a linear
combination of SSPs with fixed ages and metallicities and SSP weights taken as free
parameters. In these cases the star formation history can be in principle
derived from the observed galaxy spectra, but that requires very high signal-to-noise 
spectra \citep{steckmap}. Indeed, insufficient signal-to-noise level of
spectra leads to degeneracy between weights of the different SSPs and
unreliable star formation history. Due to understanding this effect, the majority 
of studies where linear combination of SSPs are used provide only mass- and/or 
light-weighted SSP-equivalent parameters of stellar populations that correspond to
\textsc{NBursts} fitting parameters by definition.

\subsection{Final choice of the stellar population parameters}

\begin{figure}
\centerline{
    \includegraphics[width=0.27\textwidth]{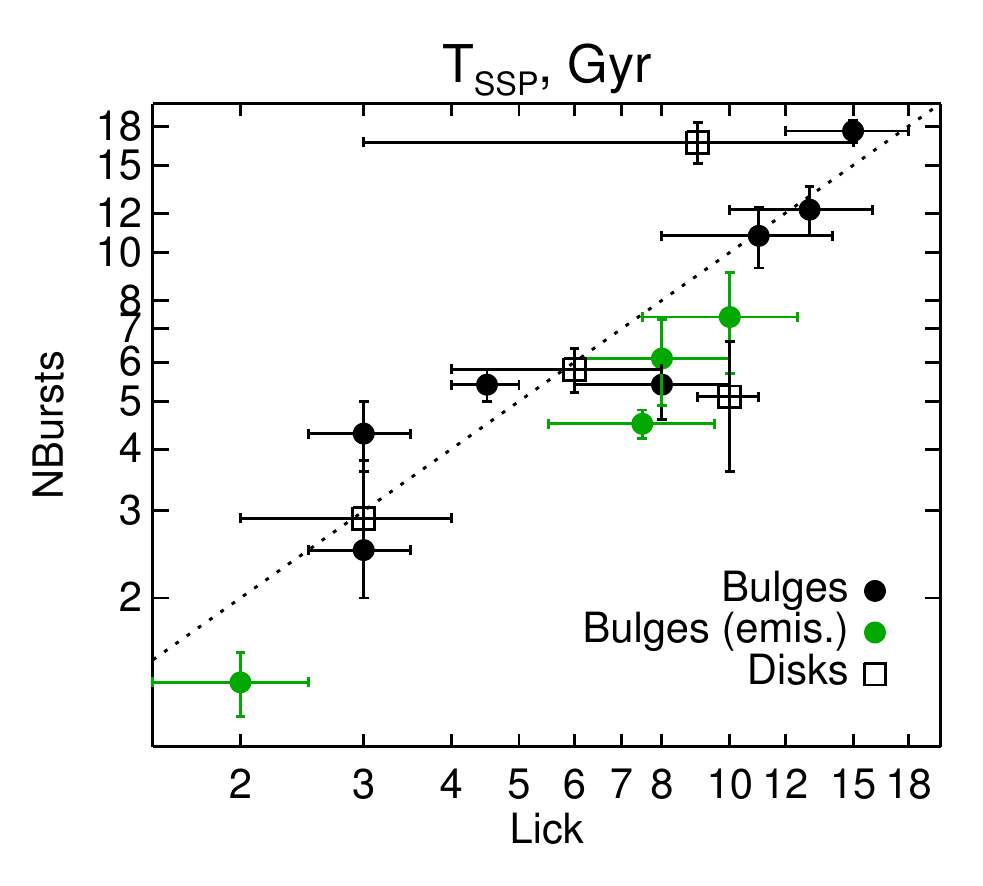}
    \includegraphics[width=0.27\textwidth]{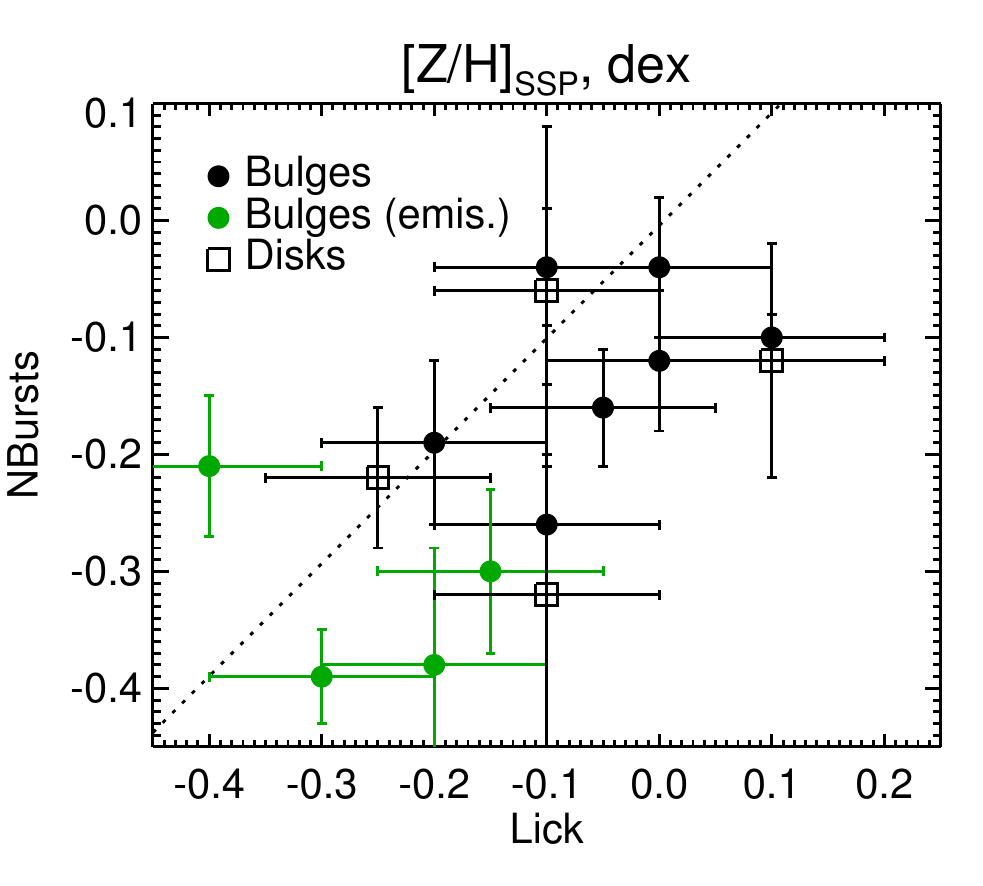}
 }
%\plottwo{katkov_fig2a}{katkov_fig2b}
\caption{The comparison of the ages and metallicities derived by two different
methods: by full spectral fitting using the PEGASE stellar population models
(`NBursts')  and through the Lick indices H$\beta$ and [MgFe] using the models
by \citet{Thomasstpop}. The dotted line represents the equality locus. Filled
circles correspond to measurements in the bulges, open squares - to disks.
Green circles show bulges with presence of emissions in the spectra.}
\label{stpopcomparison}
\end{figure}

To derive the stellar population parameters, we have tried both approaches:
we have fitted `pixel-by-pixel' all our spectra along the slit and we have
calculated the Lick indices H$\beta$, Mgb, Fe5270, and Fe5335. The full spectral
fitting included the use of the evolutionary synthesis code \textsc{pegase.hr} 
\citep{pegasehr}. The measured Lick indices were confronted to the models
by \citet{Thomasstpop} allowing to analyze magnesium-overabundant
stellar populations. To obtain the stellar magnesium-to-iron ratios, we were restricted
to the use of Lick indices only, because the full spectral fitting assumed solar abundance
ratios. On the other hand, the full spectral fitting has great advantage in
deriving the stellar population ages when we deal with the spectra containing strong 
Balmer emission lines: when the Lick index H$\beta$ is contaminated by the hydrogen
emission line, the full spectral fitting is much more safe because it allows to exclude 
spectral ranges polluted by emission lines. By calculating the Lick index H$\beta$, 
we tried to correct it for the emission by applying our approach basing on the 
measurements of the H$\alpha$ emission-line equivalent width \citep{sil2006}; 
however when the emission is strong, and the ionized-gas excitation is uncertain, 
the correction cannot be perfect. Unfortunately,
among our southern sample of the isolated S0s, almost all galaxies demonstrate
rich emission-line spectra. So for the present sample in particular we shall
analyze mostly the results on the ages and metallicities obtained through the full
spectral fitting. However, we wonder to know if two approaches give consistent
results, and for a few our galaxies with weak or absent emission lines
we have compared the ages and metallicities of the bulges and 
disks obtained by full spectral fitting using the PEGASE stellar population models 
with those obtained through the Lick indices H$\beta$ and [MgFe] using the models 
by \citet{Thomasstpop}. To get a sufficient statistical level of the comparison,  
we have involved the results for 
our previous sample of the northern isolated S0s \citep{katkov_ilg_stpop}, and the final
comparison can be inspected in Fig.~\ref{stpopcomparison}. The ages are consistent 
within the accuracy of their determination, and the metallicities may have a systematic 
shift by some 0.1 dex, perhaps due to slightly non-solar magnesium-to-iron ratios of
the stellar populations in the S0s studied by us.  

% Figure 2 %%%%%%%%%%%%%%%%%%%%%%%%%%%%%%%%%%%%%%%%%%%%%%%%%%%%%%%%%%%%%%%%%%%%%%%%%%%%%%%%%%%%%%%%%%%

\subsection{Ionized gas}

A warm-gas emission-line spectrum can be obtained by subtracting the stellar component
contribution (i.e., the best-fitting stellar population model) from the
observed spectrum at every spatial bin. The resulted pure emission-line spectra are
uncontaminated by absorption lines of the stellar components that is especially
important for the Balmer lines. Then we fitted emission lines with Gaussians
pre-convolved with the instrumental LSF in order to determine the LOS
velocities of the ionized gas and emission-line fluxes.

\section{Results}\label{txt:results}

By applying the above-mentioned techniques to every galaxy spectrum along
the major axis, we have derived the radial profiles of the following
characteristics: stellar rotation velocity, stellar velocity dispersion,
SSP-equivalent stellar ages, metallicities, and magnesium-to-iron ratios,
ionized-gas rotation velocity, warm-gas velocity dispersion, emission-line flux
ratios.  The latter characteristics can be plotted at the classical
excitation-diagnostic diagrams, so called BPT (after Baldwin, Phillips, \& Terlevich) 
\citep{BPTdiag}, to identify the gas excitation mechanism.  If we see that the
gas is ionized by young stars, we can apply so called `strong-line
calibrations' to estimate the gas metallicity. We have explored the formulae
from the paper by \citet{PPcalib} and have estimated oxygen abundance using
emission-line indices
$O3N2=\log_{10}\left(([O\iii]\lambda5007/H\beta)/([N\ii]\lambda6583/H\alpha)\right)$
and $N2=\log_{10}([N\ii]\lambda6583/H\alpha)$ which are calibrated by
\citet{PPcalib} against $12+\log (O/H)$ by using the data on 137 H\ii-regions
with known electronic temperatures.  In the case of very noisy \Hb and [O\iii]
emission lines we used another calibration from \citet{PPcalib} that involves
only $N2$ index.  Due to exploration of ratios of nearest emission lines the
dust attenuation does no affect the oxygen abundance estimations. 

Below we present the results for every galaxy in graphical way and give brief
description of the individual properties of the galaxies.

\noindent
{\bf IC 1608.} It is one of the most luminous galaxies of our sample. The
galaxy demonstrates fast stellar rotation; however the stellar disk is rather
hot dynamically, $\sigma _* \ge 100$ \kms.  The galaxy is very gas-rich; we
observe strong emission lines over the full slit extension.  The nuclear
emission-line spectrum is a typical LINER-like one; starting from the radius of
about 10 arcsec, the ionized gas is excited mostly by young stars, and in the
outer part, $R\approx 40$\arcsec, a starforming ring is clearly seen. The gas
subsystem looks cold only within the starforming ring; elsewhere in the disk
the gas velocity dispersion matches that of stars. The gas rotation curve
coincides exactly with the stellar one so we can be sure that the gas is
confined to the main galaxy plane. The gas oxygen abundance at $R>30\arcsec$ is
observed to be nearly solar or slightly higher while the stellar mean
metallicity there is very poor, about $-0.6\div -0.9$ dex. The
magnesium-to-iron ratio in the stellar component is homogeneous along the
radius and close to $+0.1$ dex; the mean stellar age looks intermediate, 3--5
Gyr, over the whole galaxy too. We may so suspect that low-level starforming
events like the current one have occured multiple times during the galaxy
evolution for the last several billion years; perhaps they have been provoked
by small gas-rich satellites merging.\\

\noindent
{\bf IC 4653.} This dwarf galaxy is classified as SB0/a pec in the
NED\footnote{NASA/IPAC Extragalactic Database} database. However despite this
relatively late morphological type and the elongated isophote shape, our
spectrograph slit aligned with the isophote major axis reveals very weak
rotation and rather large stellar velocity dispersion.  We would re-classified
the galaxy as a dwarf elliptical one and exclude it from the further
consideration of our sample of isolated lenticular galaxies. Interestingly, the
strong emission lines excited by the current star formation are seen over the
whole galaxy including its nucleus.

\noindent
{\bf NGC 1211.} This luminous, almost face-on galaxy has a bar and two rings --
the inner reddish one at $R\approx 20\arcsec$ and the blue starforming outer
one at $R\approx 60\arcsec$. Surveys in the 21~cm line reported earlier large
neutral hydrogen content in this galaxy, 5.5 billion solar masses of HI
\citep{Garcia-Appadoo_2009}, so the strong emission lines were also expected in
its optical spectrum.  The discrepancy between the rotation velocities of the
ionized gas and the stars at $R<10\arcsec$ can be explained both by asymmetric
drift and bar influence; beyond the radius of 10\arcsec\ the ionized gas and
stars rotate together, and we conclude that the gas is confined to the main
galaxy plane. The gas is excited by the LINER-like nucleus and by shock waves
(from the bar?) in the central part of the galaxy, at $R<10\arcsec$, and by
young stars beyond this radius; the oxygen abundance of the gas in the outer
starforming ring is the solar one. Meantime the low-surface brightness stellar
disk at $R\approx 30\arcsec$ demonstrates very old age, $T\ge 10$~Gyr, and very
low stellar metallicity, [Z/H]$\approx -1.5$ dex. The bulge and the lens which
we relate to the bar ends, demonstrate intermediate stellar ages and strong
metallicity gradient along the radius.

\noindent
{\bf NGC 2917.} This very luminous S0$^+$ galaxy is strongly inclined to the
line of sight, but is not exactly edge-on, so we can distinguish a dust ring
and no signs of bar in the galaxy. The bulge is so small that the galaxy has
been included into the list of `flat' late-type galaxies by
\citet{Mitronova_2004}. The systemic velocity of NGC 2917 given in the NED,
3675 \kms, is erroneous coming from a very weak spurious 21-cm signal detected
by \citet{Richter_1987}. Our optical spectral observations give $v_{sys}=5377$
\kms\ for this galaxy, so the galaxy is even more luminous than it has been
thought before. Though we have traced the stellar component of NGC~2917 almost
up to its optical border, $R_{25}=38\arcsec$ (RC3\footnote{Third Reference
Catalogue of Bright Galaxies.}), we have only measured its lens; the outer
stellar disk is very low-surface brightness one and could not be detected in
our observations. While the bulge has intermediate stellar-population
characteristics, the lens looks rather young, $T=2-3$~Gyr, that is consistent
with the ionized gas excited by current star formation at $R\ge 10\arcsec$.
The ionized-gas metallicity is high. The fall of the ionized-gas rotation
velocity at the southern edge of the galaxy accompanied by the rise of the
velocity dispersion of the gas clouds seems to be real. Are there any traces of
interaction?

\noindent
{\bf NGC 3375.} Another galaxy which being previously classified as a
lenticular one is in fact an elliptical: its stellar component does not rotate
regularly, and the stellar velocity dispersion exceeds 150 \kms\ everywhere
through the galaxy. Emission lines are absent in the spectrum.

\noindent
{\bf NGC 4240.} The galaxy is classified in RC3 as between E and S0 ($T=-3.8\pm
0.5$). In the frame of the APM survey \citep{naim_1995} when 6 independent
researchers classified it `by eye', three voted for S0 and three voted for E
(see `Detailed classification' option in the NED). However our long-slit
cross-section along the major axis reveals rather fast rotation of the stellar
component; and in the photometric data we see an exponential disk at
$R>15\arcsec$. So we consider NGC~4240 as a lenticular galaxy. The stellar
population properties were spectrally studied by \citet{Reda_2007} up to the
distance about 10\arcsec\ from the center; a slightly subsolar metallicity and
a rather old age were measured by the Lick index method. The kinematics was
examined by \citet{Hau_2006}; however their slit was obviously off the
dynamical center, and they did not report gas counterrotation for this galaxy
which is striking. We have traced the stellar rotation and stellar population
properties toward $R\approx 25\arcsec$ so measuring not only the bulge but also
the large-scale stellar disk at $R>15\arcsec$. Both the bulge and the disk have
an intermediate stellar age, about 5~Gyr, but the disk is very metal-poor,
[Z/H]$\approx -1.0$, while the bulge has only [Z/H]$\approx -0.3$. The ionized
gas counterrotates the stars in the bulge-dominated area; while the stellar
velocity dispersions of the stars and gas clouds are comparable, the rotation
velocities differ significantly, and we conclude that the ionized gas may
rotate in the plane which does not coincide with the plane of the stellar disk.
The gas metallicity in the outer starforming ring, at $R\approx 15\arcsec$, is
close to the solar one.

\noindent
{\bf NGC 6010.} It is another small-bulge, edge-on S0 galaxy included into the
catalogue of `flat galaxies' by \citet{Mitronova_2004}. Also we must note that
in our present sample it is the only S0-galaxy without strong emission lines in
the spectra. We see only weak narrow emission lines with LINER-like excitation
in the very circumnuclear region; some signs of the ionized-gas counterrotation
are however detected. Meantime the neutral hydrogen is found in this galaxy by
\citet{Springob_2005}, but no signs of recent star formation are present. The
stellar ages of the nucleus ($T=9$~Gyr) and of the bulge ($T=8$~Gyr) are
slightly older than in other galaxies of our sample. However, the
characteristics of the disk in the radius range $R=20\arcsec - 40\arcsec$ are
quite typical -- $T=5$~Gyr and [Z/H]$=-0.4$.

\noindent
{\bf NGC 7693.} Due to instrumental problems, the galaxy was observed with the
slit turned by some 40 degree to the major axis. However, even so, the observed
stellar rotation is too slow, and the ionized-gas velocities are quite
decoupled from the stellar ones. Consequently, no signs of current star
formation is seen in this galaxy, though both the bulge and the disk look very
young, 1--3~Gyr old.  The magnesium-to-iron ratio over the whole galaxy is
solar so it seems that continuous star formation ceased rather recently, due to
perhaps just minor merger from an inclined orbit.

\noindent
{\bf UGC 9980.} The galaxy demonstrates fast regular rotation, looking quite
similar in the stellar and ionized-gas components. The gas is spread over the
whole galaxy, and starting outward from the radius $R > 10\arcsec$ it is
excited by young stars. However, both the bulge and the large-scale stellar
disk possess rather old stellar populations, $T=7$~Gyr in the former and
$T=10$~Gyr in the latter, so the widespread star formation has evidently
started quite recently: unless the case of NGC 7693, this time minor merging
has stimulated star formation, not ceased it. The difference of metallicities
-- [Z/H]$=-1.0$ dex in the stellar disk and --0.2 dex in the gaseous disk --
indicates also the external origin of the current fuel for the star formation.
The inner stellar ring related to ansae at the ends of the bar, at $R\approx
10\arcsec$, is distinguished by slightly younger stellar age, $T\approx 5$~Gyr.
However, this ancient ring-like star formation burst was probably related not
to interaction but to the bar affecting gaseous disk of the galaxy which was
perhaps more gas-rich at $z=0.5$ than it is at the present epoch. 

\medskip

After obtaining the full radial profiles of the stellar characteristics in the
galaxies studied, which are presented in the Appendix, we have wished to
extract mean  characteristics for the large-scale galaxy components -- bulges
and disks.   To identify radius ranges that correspond to the bulge and disk
domination in the integrated light, we have undertaken photometric
decomposition of the images of the galaxies. For this purpose we have used
mostly the SDSS public database, Data Release 9; the $r$-band images as the
images with the highest signal-to-noise ratio have been taken. For the
southern galaxy NGC~4240 which was not observed in the frame of the SDSS we
have decomposed the 2MASS composite, $J+H+K$, image.  For one galaxy, IC~1608,
very deep $gri$ photometric data obtained during test observations of the
LCOGT project \citep{sil_lcogt}. For every galaxy, we have performed an
isophotal analysis and have derived azimuthally-averaged surface-brightness
radial profiles.  By inspecting these profiles, we have found the outer radial
segments where the surface-brightness radial profiles can be well approximated
by exponential laws, and the isophote ellipticity stays constant. These outer
parts of galaxies are identified by us as disk-dominated. To characterize the
bulges which are mostly compact in the galaxies of our sample we fix the
radial range of $4\arcsec -7\arcsec$ that is beyond the unresolved nucleus
contamination under our seeing conditions. In some galaxies we have also
distinguished the radial ranges where we see rings of enhanced stellar
brightness or flat brightness profile segments betraying the presence of
lenses. The corresponding segments for each component are shown by
shaded gray lines in figures in the Appendix.  The mean stellar ages,
metallicities, and magnesium-to-iron ratios for the bulges, disks, rings, and
lenses of the galaxies studied here are presented in the
Table~\ref{tbl_stpop}.

% Table 3 %%%%%%%%%%%%%%%%%%%%%%%%%%%%%%%%%%%%%%%%%%%%%%%%%%%%%%%%%%%%%%%%%%%%%%%%%%%%%%%%%%%%
\begin{deluxetable*}{lrrrrr}
\tablecolumns{6}
\tablewidth{0pc}

\tablecaption{Average stellar population parameters.\label{tbl_stpop}}

\tablehead{
\colhead{Galaxy} & \colhead{N(bins)} & \colhead{T, Gyr} & \colhead{[Z/H], dex} & \colhead{[Mg/Fe], dex} & 
\colhead{$\sigma$, km/s} 
}

\tabletypesize{\scriptsize}

\startdata
%\cutinhead{Bulge}
\multicolumn{6}{c}{Bulge}\\
\hline
IC 1608 &      4 & $    4.7^{\pm 0.3}$ & $   -0.21^{\pm 0.07}$ & $     0.12^{\pm  0.09}$ & $  149^{\pm  9}$ \\
%IC 3152 &      4 & $    5.2^{\pm 0.14}$ & $   -0.27^{\pm 0.06}$ & $     0.12^{\pm 0.09}$ & $  187^{\pm 32}$ \\
%IC 4653 &      6 & $    1.6^{\pm 0.3}$ & $   -1.02^{\pm 0.06}$  &        \nodata              & $   136^{\pm 21}$ \\
NGC 1211 &      3 & $    4.5^{\pm 0.4}$ & $   -0.16^{\pm 0.05}$ & $     0.11^{\pm  0.07}$ & $  156^{\pm 18}$ \\
NGC 2917 &      4 & $    6.1^{\pm 1.3}$ & $   -0.21^{\pm 0.06}$ & $     0.27^{\pm  0.08}$ & $  191^{\pm  4}$ \\
%NGC 3375 &      3 & $    4.5^{\pm 0.3}$ & $   -0.39^{\pm 0.04}$ & $     0.18^{\pm  0.06}$ & $  159^{\pm  2}$ \\
NGC 4240 &      4 & $    4.6^{\pm 0.3}$ & $   -0.32^{\pm 0.08}$ & $     0.18^{\pm  0.09}$ & $  108^{\pm  11}$ \\
NGC 6010 &      4 & $    8.16^{\pm 0.45}$ & $   -0.19^{\pm 0.07}$ & $     0.19^{\pm  0.06}$ & $  154^{\pm  11}$ \\
NGC 7693 &      8 & $    1.35^{\pm 0.18}$ & $   -0.38^{\pm 0.10}$ & $    -0.02^{\pm  0.02}$ & $   82^{\pm 14}$ \\
UGC 9980 &      6 & $    7.4^{\pm 1.4}$ & $   -0.30^{\pm 0.07}$ & $     0.18^{\pm  0.11}$ & $  138^{\pm 20}$ \\
\hline
\multicolumn{6}{c}{Disk}\\
\hline
IC 1608 &      8 & $    3.5^{\pm 0.8}$ & $   -0.46^{\pm 0.14}$ & $     0.18^{\pm  0.15}$ & $  138^{\pm 32}$ \\
%IC 3152 &      7 & $    3.9^{\pm 1.7}$ & $   -0.98^{\pm 0.16}$ & $     0.20^{\pm  0.13}$ & $  172^{\pm 4}$ \\
%IC 4653 &     15  & $  1.4^{\pm 0.4}$ & $   -0.78^{\pm 0.23}$ &   \nodata                    & $  118^{\pm 36}$ \\
NGC 1211 &      2 & $   10.5^{\pm 4.1}$ & $   -1.50^{\pm 0.14}$ & \nodata                     & $  141^{\pm 12}$ \\
NGC 2917 &      0 & \nodata & \nodata & \nodata                      & \nodata \\
%NGC 3375 &      0 & \nodata & \nodata & \nodata                      & \nodata \\
NGC 4240 &      5 & $    5.4^{\pm 2.1}$ & $   -1.02^{\pm 0.11}$ & $     0.33^{\pm  0.13}$ & $  112^{\pm 8}$ \\
NGC 6010 &     12 & $    5.4^{\pm 2.4}$ & $   -0.36^{\pm 0.16}$ & $     0.18^{\pm  0.04}$ & $  113^{\pm 21}$ \\
NGC 7693 &     11 & $    1.5^{\pm 0.9}$ & $   -0.67^{\pm 0.21}$ & $     0.15^{\pm  0.13}$ & $  106^{\pm 24}$ \\
UGC 9980 &      5 & $    9.8^{\pm 2.8}$ & $   -0.99^{\pm 0.12}$ & $     0.21^{\pm  0.20}$ & $   83^{\pm 5}$ \\
\hline
\multicolumn{6}{c}{Lens/Ring/Plateu}\\
\hline
IC 1608 &     10 & $     4.6^{\pm 2.8}$ & $   -0.77^{\pm 0.20}$ & $     0.24^{\pm  0.08}$ & $   98^{\pm 5}$ \\
%IC 3152 &      0 & \nodata & \nodata & \nodata & \nodata \\
%IC 4653 &      0 & \nodata & \nodata & \nodata & \nodata \\
NGC 1211 &     12 & $    5.6^{\pm 2.7}$ & $   -0.82^{\pm 0.23}$ & $     0.20^{\pm  0.18}$ & $  148^{\pm 38}$ \\
NGC 2917 &     10 & $    2.6^{\pm 0.6}$ & $   -0.34^{\pm 0.08}$ & $     0.24^{\pm  0.07}$ & $  130^{\pm 21}$ \\
%NGC 3375 &     12 & $    4.9^{\pm 1.6}$ & $   -0.61^{\pm 0.18}$ & $      0.06^{\pm  0.11}$ & $  160^{\pm 12}$ \\
NGC 4240 &      0 & \nodata & \nodata & \nodata & \nodata \\
NGC 6010 &      0 & \nodata & \nodata & \nodata & \nodata \\
NGC 7693 &      0 & \nodata & \nodata & \nodata & \nodata \\
UGC 9980 &      7 & $     4.6^{\pm  2.2}$ & $     -0.48^{\pm  0.18}$ & $  0.22^{\pm  0.19}$ & $     114^{\pm 8}$ \\
\enddata
\end{deluxetable*}

\section{Discussion}\label{txt:discus}

\begin{figure*}

\plottwo{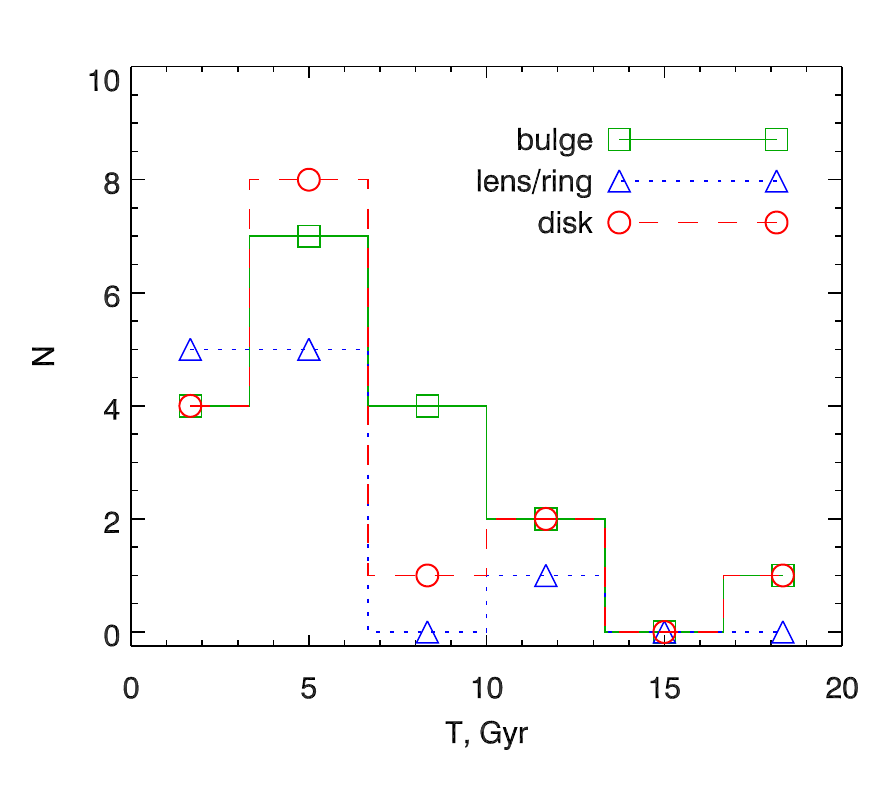}{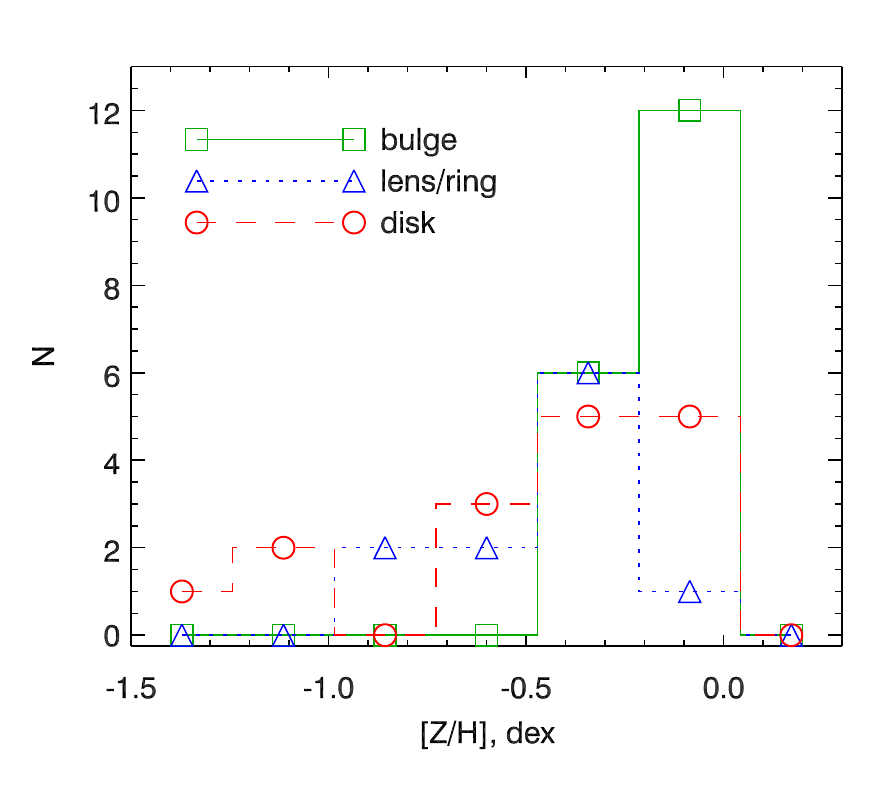}
\plottwo{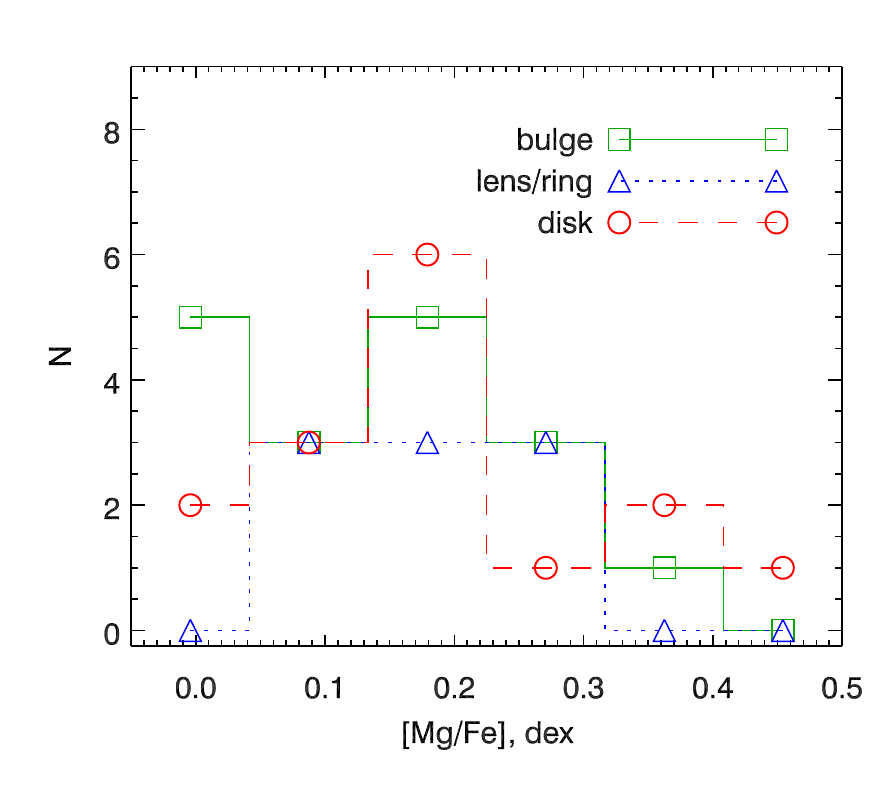}{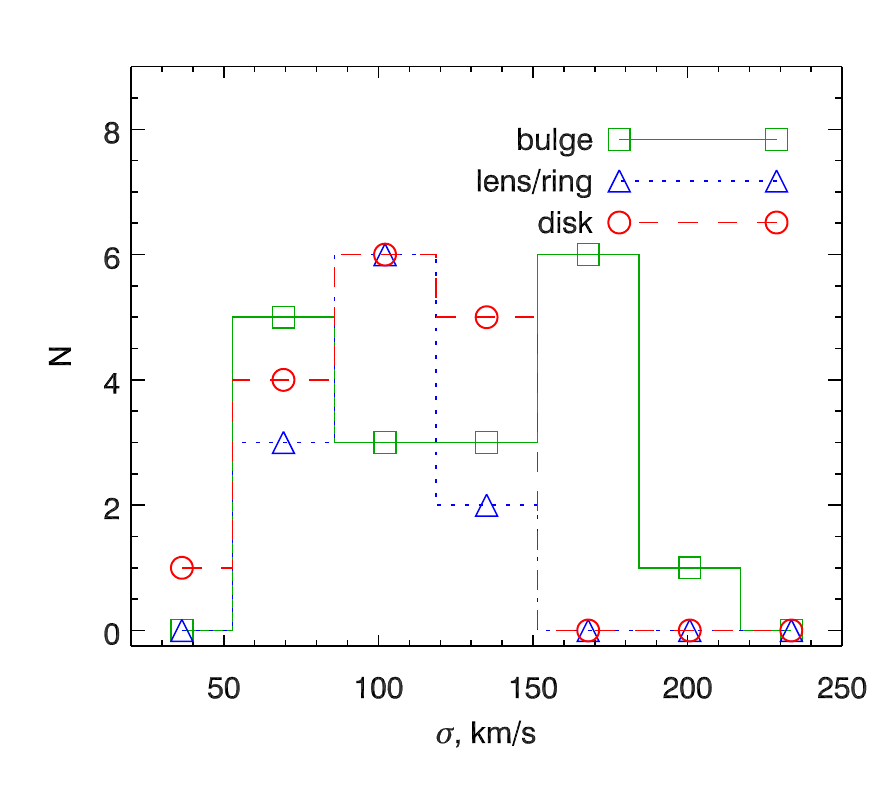}
\caption{Distributions of our complete sample of the isolated lenticular
galaxies over the found parameters of the stellar populations for every
structural component -- bulges, disks, rings or lenses.}
\label{allhist}
\end{figure*}

\begin{figure}
\plotone{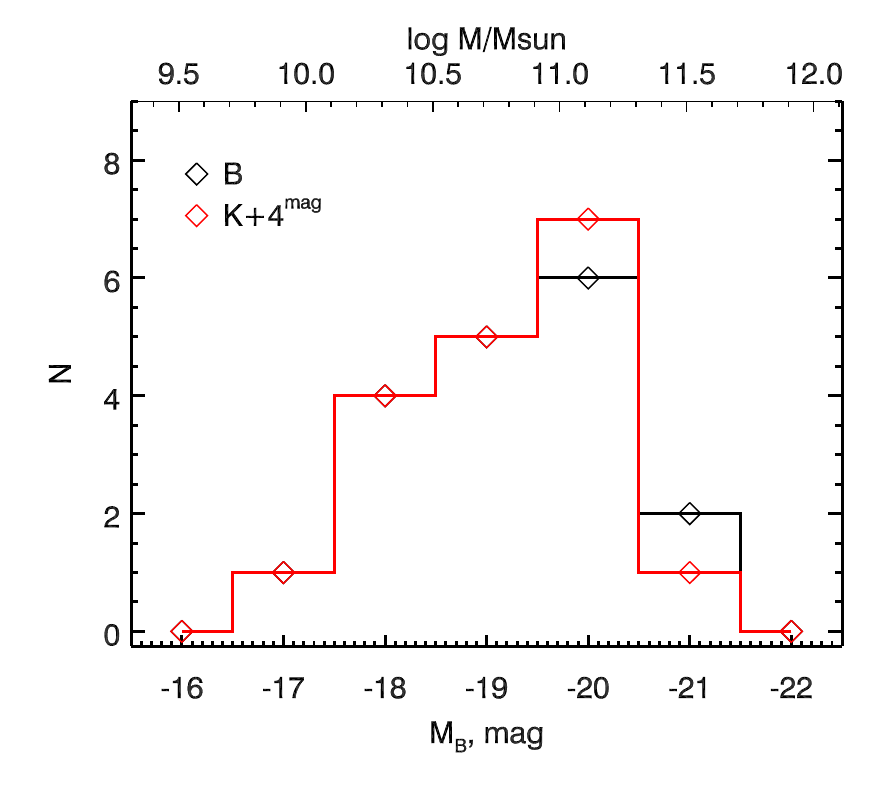}
\caption{Histogram of integrated absolute magnitudes in B-band (black line) and shifted by
$4^m$ K-band (red line) for the total SALT \& SCORPIO sample of the isolated lenticular
galaxies. Corresponding stellar masses are shown in the upper \textit{x}-axis and
were calculated by using $M/L_K=1$.}
\label{mabshist}
\end{figure}

In this paper we have described the results of spectral study for 7 isolated
lenticular galaxies of the southern sky (another 2 galaxies observed so far
have been re-classified here as ellipticals basing on their stellar kinematics and
do not take part in the analysis below). Earlier we have already published the
results of the similar study for 11 isolated lenticulars of the northern sky
which were observed at the Russian 6m telescope using the SCORPIO and SCORPIO-2 spectrograph
\citep{ilg_gas, katkov_ilg_stpop}. With the totality of 18 isolated lenticular
galaxies observed with the long-slit spectrographs of two large telescopes, we
can now discuss some statistical properties concerning the kinematics, the stellar
population parameters, and the ionized-gas features in isolated lenticular
galaxies. The overall distributions of the parameters of the stellar component
for the bulges, disks, and rings/lenses are presented for the total sample in
Fig.~\ref{allhist}. The distribution of the absolute magnitudes in $B$- and $K$-band are shown
in Fig.~\ref{mabshist}.

% Figure 3 %%%%%%%%%%%%%%%%%%%%%%%%%%%%%%%%%%%%%%%%%%%%%%%%%%%%%%%%%%%%%%%%%%%%%%%%%%%%%%%%%%%%%%%%%%%

% Figure 4 %%%%%%%%%%%%%%%%%%%%%%%%%%%%%%%%%%%%%%%%%%%%%%%%%%%%%%%%%%%%%%%%%%%%%%%%%%%%%%%%%%%%%%%%%%%

\subsection{Bulges vs. Disks}

\begin{figure*}
\centerline{
    \includegraphics[width=0.33\textwidth]{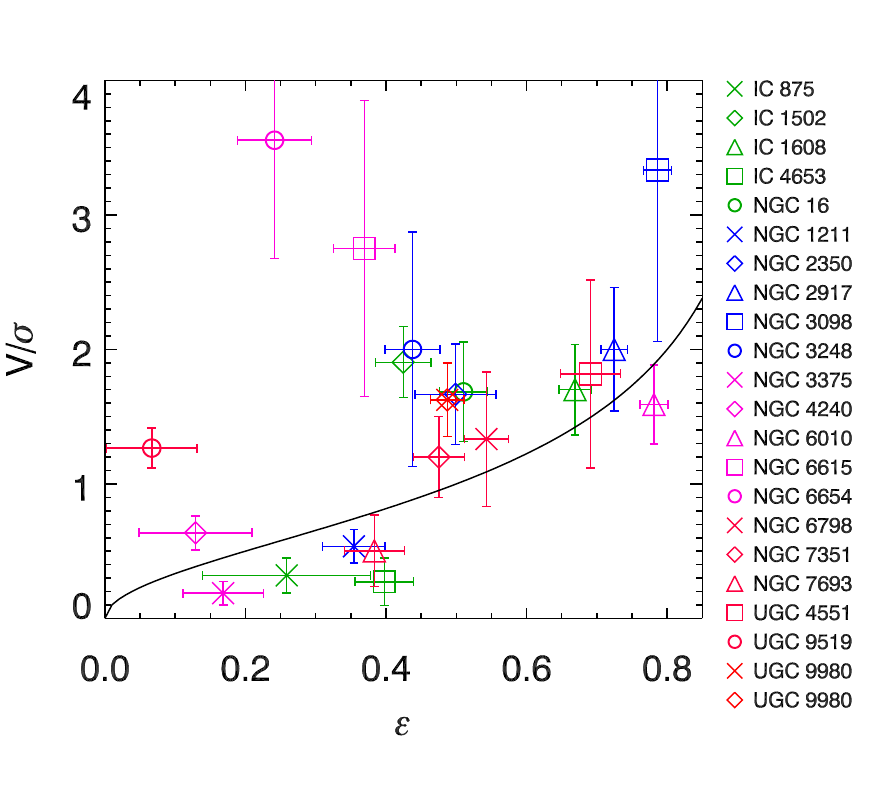}
    \includegraphics[width=0.33\textwidth]{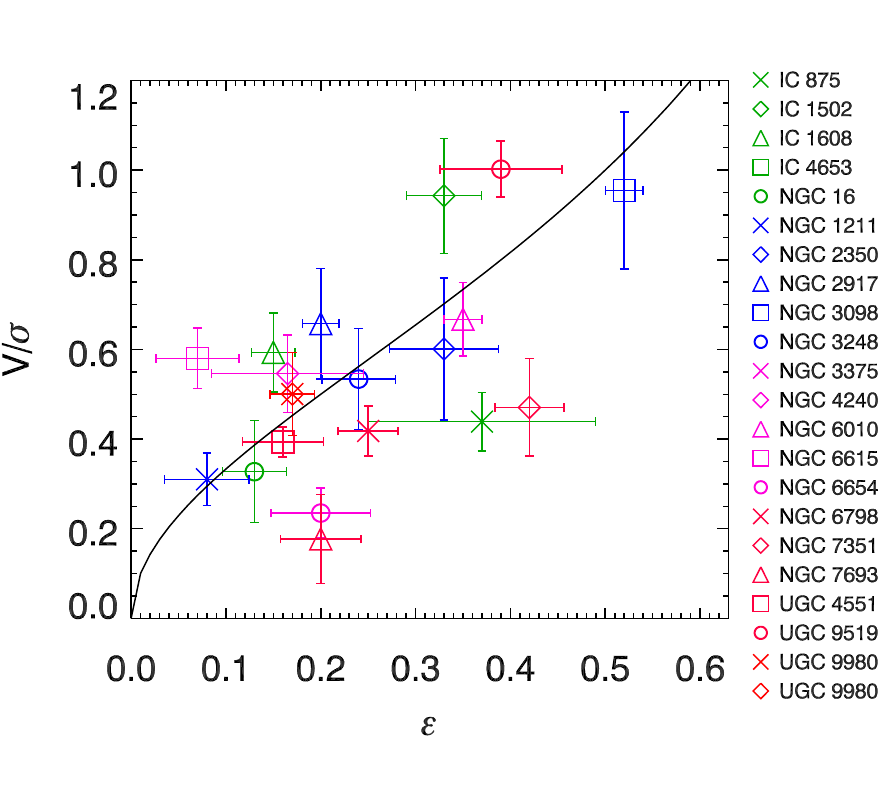}
    \includegraphics[width=0.33\textwidth]{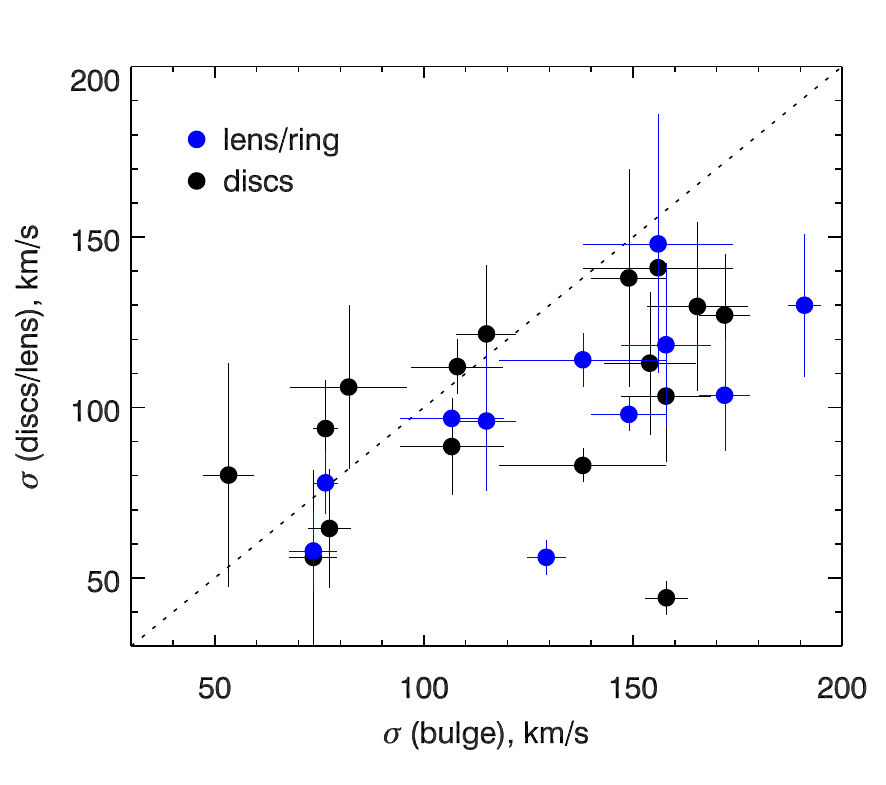}}

\caption{Dynamical diagnostics diagrams: ellipticity vs. $V/\sigma$ for the
disks (\textit{left}), the same for the bulges (\textit{center}), and
comparison of the stellar velocity dispersions in the bulges and disks
(\textit{right}). The solid line shows the locus of oblate spheroids,
which can be considered as demarcation line between disk-like and 
spheroid-like system. }
\label{binneydia}   
\end{figure*}

Fig.~\ref{binneydia} demonstrates first of all the dynamical status of the
bulges and disks in our sample of isolated lenticular galaxies. The diagrams
presented in the two left plots for the disks and for the bulges
correspondingly, confront the ratio of the regular rotation velocity to the
stellar velocity dispersion versus the visible ellipticity of the isophotes. It
was proposed by \citet{Illingworth_1977} and theoretically calculated by
\citet{Binney_1978_rotellip,Binney_1978_prol} to check if the shape of a
galaxy spheroid is supported by rotation. The main theoretical locus at this
diagram signifies so called oblate spheroids -- ones round in the equatorial
plane, with isotropic velocity dispersion, whose smaller third axis is
completely explained through flattening by rotation. Many true elliptical galaxies
are found well below this locus because they rotate too slowly, and their
shapes are supported by anisotropy of the velocity dispersion distributions.
Jonh \citet{Kormendy_1993} (see also \citet{Kormendy_Kennicutt_2004} for the
updated version of this diagram) used this diagram for the bulges of disk
galaxies to separate so called `classical bulges' which can be considered as
analogues of elliptical galaxies, from the `pseudobulges' which reveal the
disk-like kinematics. If the observed characteristics place some bulges above
the theoretical line for oblate spheroids, we would classify them as
`pseudobulges' formed from the disk material during secular dynamical
evolution. Fig.~\ref{binneydia}, middle plot, gives evidence for the roughly equal
proportion of `classical bulges' and `pseudobulges' among the isolated
lenticular galaxies: the points are oscillating around the theoretical locus
for the oblate isotropic spheroids.  When inspecting Fig.~\ref{binneydia}
(right plot), we make sure again that in 8-10 bulges of 18 the stellar velocity
dispersion is the same as in the surrounding disks so indeed these are
`pseudobulges'.

% Figure 5 %%%%%%%%%%%%%%%%%%%%%%%%%%%%%%%%%%%%%%%%%%%%%%%%%%%%%%%%%%%%%%%%%%%%%%%%%%%%%%%%%%%%%%%%%%%

\begin{figure*}
\centerline{
    \includegraphics[width=0.33\textwidth]{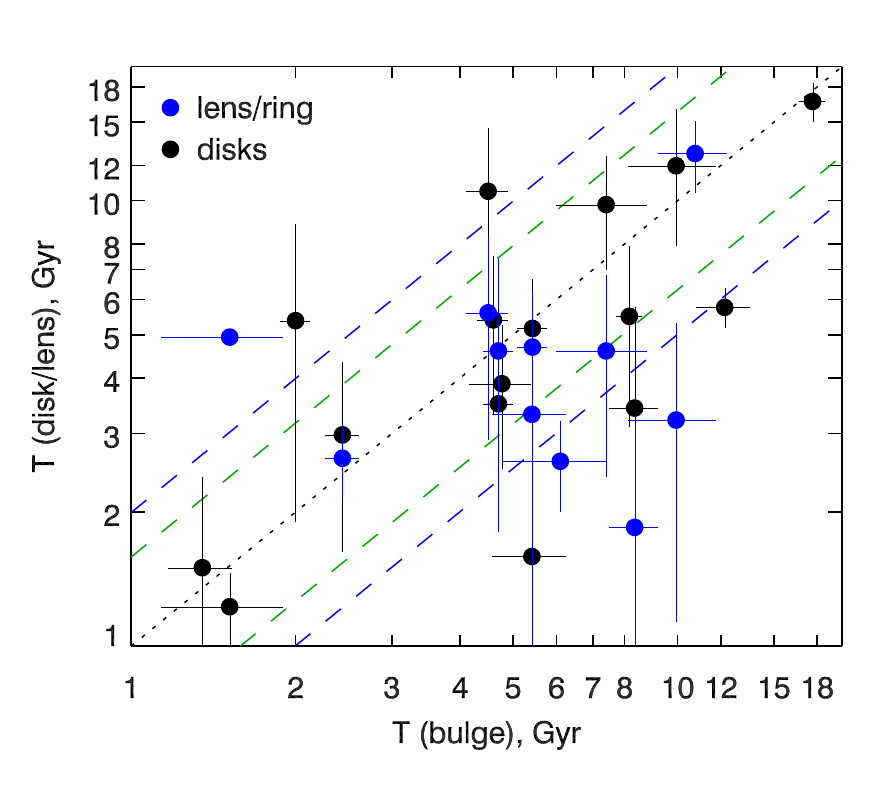}
    \includegraphics[width=0.33\textwidth]{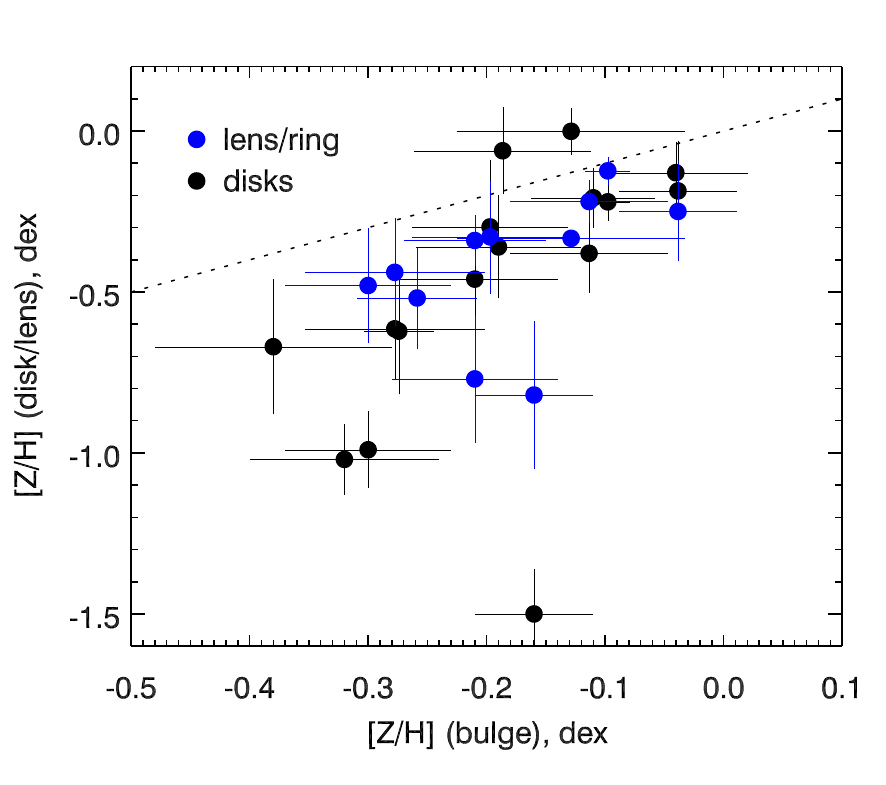}
    \includegraphics[width=0.33\textwidth]{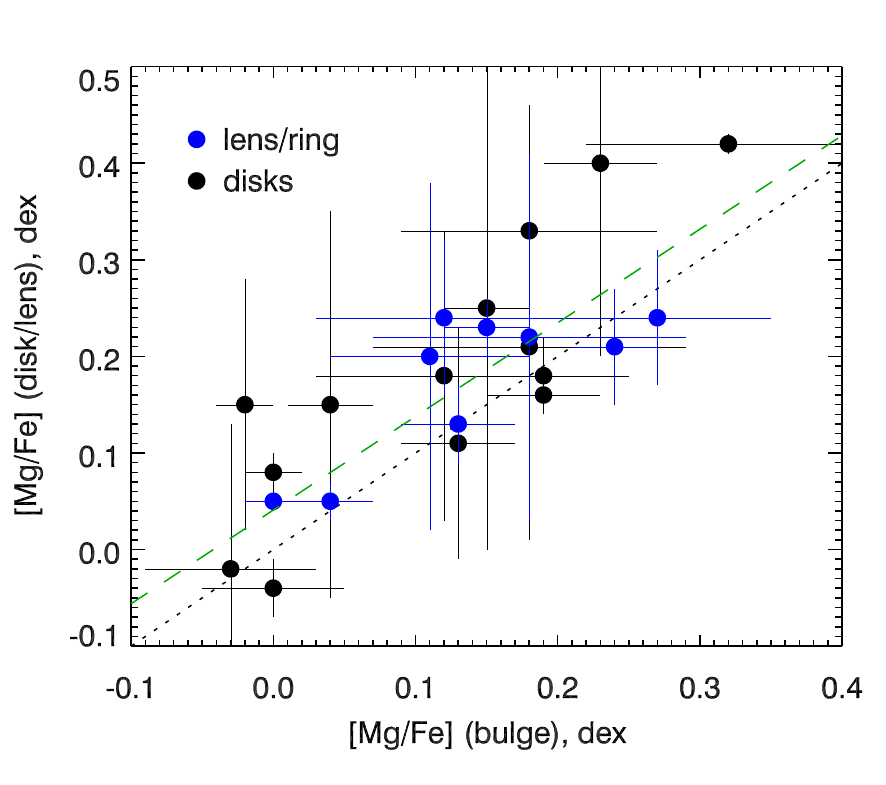}}
\caption{Comparison of the stellar populations in the bulges and in the disks:
Age--age diagram (\textit{left}), metallicity-metallicity diagram
(\textit{center}), and [Mg/Fe] - [Mg/Fe] (\textit{right}). Green and blue
dashed lines in the left panel correspond to deviation from bisector (dotted line) of value
$\pm0.2$ dex (by 1.5 times) and $\pm0.3$ dex (by 2 times), respectively.
Dotted lines show equality line. Green dashed line in the right panel shows linear fit of
measurements.}
\label{bulgedisc}
\end{figure*}

Fig.~\ref{bulgedisc} gives comparison of the characteristics of the stellar
populations in the bulges and in the disk structures: it is the comparison of
age--age, metallicity--metallicity, Mg/Fe ratio--Mg/Fe ratio.  The first and
the third plots demonstrate correlations between the properties of the bulges
and of the disks: covering all the range of possible ages, from 1~Gyr to
17~Gyr, the mean stellar ages of the bulges and disks tend to be similar in the
galaxies studied, and the magnesium-to-iron ratios are strictly the same in the
bulges and in the disk structures. It is an opposition to the properties of S0
galaxies in denser environments: \citet{sil_s0} found for a sample of mostly
group S0 members that the disks appear to be usually older than the bulges
covering homogeneously the upper left corner of the diagram similar to the
Fig.~\ref{bulgedisc} (left plot), and \citet{virgo_gemini} have found just the
same effect for all the Virgo S0s studied by them. In both dense-environment samples
the stellar disks appear to be much more magnesium-overabundant than the bulges. 
We can conclude that when placed beyond the outer
gravitational and hot-medium influence, the bulges and the disks in S0 galaxies
formed synchronously: star formation started simultaneously here and there and
ceased at one moment. Interestingly, despite the synchronous star formation,
the mean stellar metallicities of the disk structures are significantly lower
than the metallicities of the bulges (Fig.~\ref{bulgedisc}, middle plot). Does
it mean that pristine outer gas was accreted by the outer disks and fueled
star formation there, while the nearly simultaneous star formation in the bulges was
fed by the gas pre-processed and enriched in the disks?

% Figure 6 %%%%%%%%%%%%%%%%%%%%%%%%%%%%%%%%%%%%%%%%%%%%%%%%%%%%%%%%%%%%%%%%%%%%%%%%%%%%%%%%%%%%%%%%%%%

\begin{figure*}
\centerline{
    \includegraphics[width=0.33\textwidth]{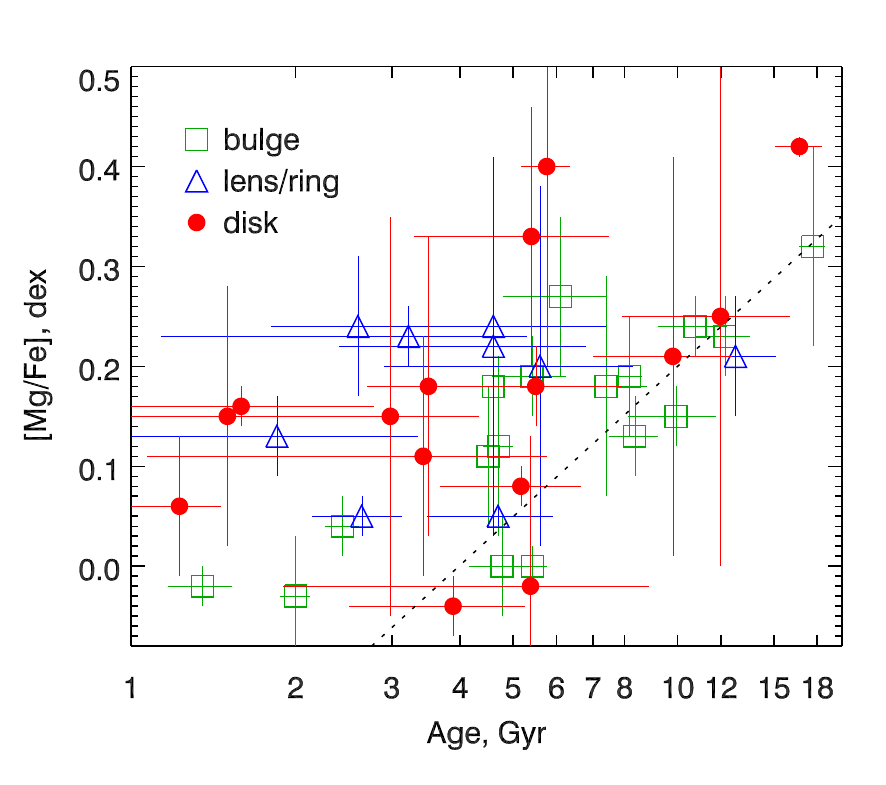}
    \includegraphics[width=0.33\textwidth]{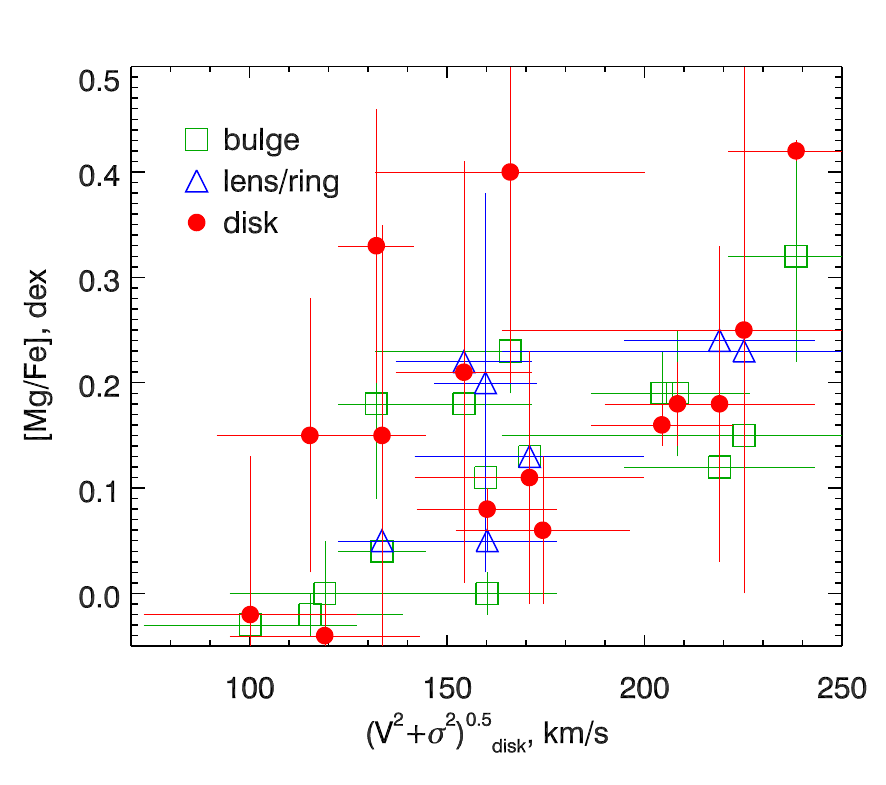}
    \includegraphics[width=0.33\textwidth]{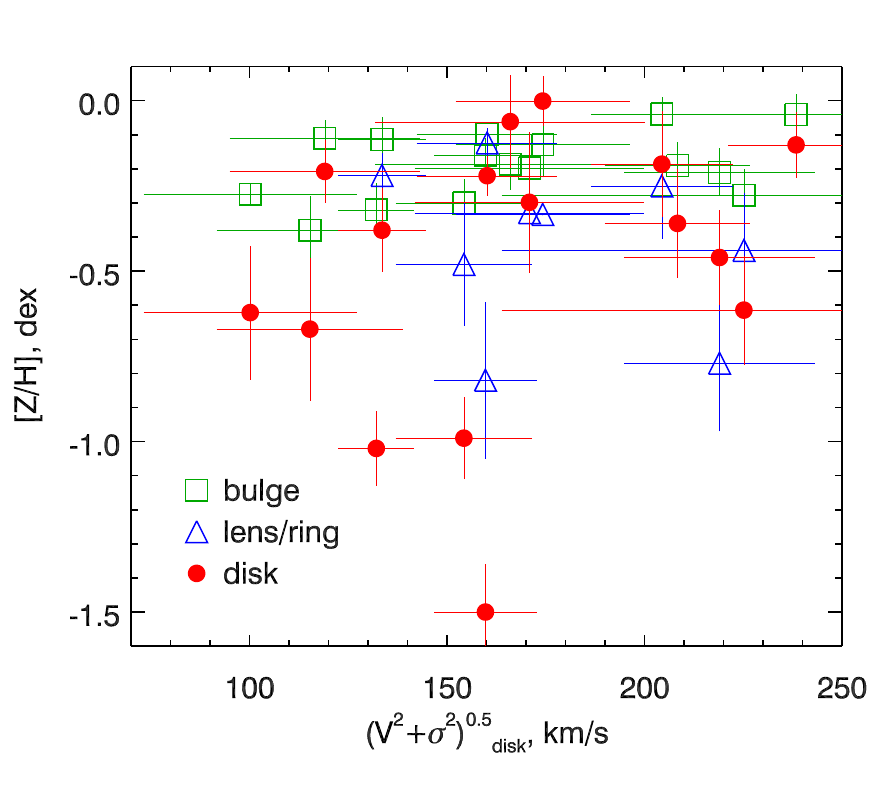}}
\caption{The relation age--Mg/Fe ratio (\textit{left}), dynamical parameter
$(v^2+\sigma^2)^{0.5}$ versus alpha-element abundance (\textit{center}),
and some kind of the mass-metallicity relation (\textit{right}).}
\label{scalrel}
\end{figure*}

Let us inspect some scaling relations connecting evolutionary and dynamical
characteristics of the stellar components which are commonly studied for the
elliptical galaxies. Fig.~\ref{scalrel} (left plot) confronts the mean stellar
ages of the different structural components with their magnesium-to-iron ratio
which characterizes the duration of the main starforming episode, from a very
brief, shorter than $10^9$~years ([Mg/Fe]$=+0.3$), to several Gigayears
([Mg/Fe]$=0.0$). We see a cloud of points limited at the down right by a linear
law which marks probably the initial epoch of launching star formation in S0s
at $z\approx 3$: star formation starting 12~Gyr ago and ceasing just
immediately would give [Mg/Fe]$=+0.3$, and star formation starting 12~Gyr ago
and lasting to 4~Gyr ago would give [Mg/Fe]$=0.0$. However there is a lot of
points, relating both to the bulges and to the disks, which are expanding to
the left of this limiting line. Obviously these are the stellar systems which
have started their formation much later than at $z=2-3$: to get the mean
stellar age of 3~Gyr and [Mg/Fe]$=+0.3$ signifying the duration of star
formation less than 1~Gyr, the process had to be launched at $z=0.4$.  From
this plot, we conclude that main star formation events both in the bulges and
in the disks of the isolated S0 galaxies have no a single fixed epoch, but are
homogeneously spread from very high redshifts to rather recent ones.

% Figure 7 %%%%%%%%%%%%%%%%%%%%%%%%%%%%%%%%%%%%%%%%%%%%%%%%%%%%%%%%%%%%%%%%%%%%%%%%%%%%%%%%%%%%%%%%%%%

Two other plots of the Fig.~\ref{scalrel} confront the chemical properties of
the stellar populations to the dynamical parameter $(v^2+\sigma ^2)^{0.5}$,
where $v$ and $\sigma$ are the rotation velocity and stellar velocity dispersion
averaged over the disk-dominated or bulge-dominated area; this dynamical parameter characterizes, 
under the condition of virialization, the local gravitational potential. The correlation of the
magnesium-to-iron ratio with the gravitational potential well, in particular, with
the central stellar velocity dispersion, is well known for the elliptical
galaxies (e.g. \citet{trager}). In our data, we see that the bulges
(spheroids) and the disks behaves similarly as concerning the duration (the
effectiveness?) of star formation in a particular gravitational potential
well: the deeper the well the shorter star formation. However, the similarity
of the bulges and disks disappears when we inspect not the Mg/Fe ratio, but
the global metallicity versus the dynamical parameter (Fig.~\ref{scalrel},
right plot): while the bulges follow the well-known mass--metallicity
relation, the larger mass the higher metallicity, this correlation vanishes
completely for the disks.

\begin{figure*}
\centerline{
\includegraphics[width=0.5\textwidth]{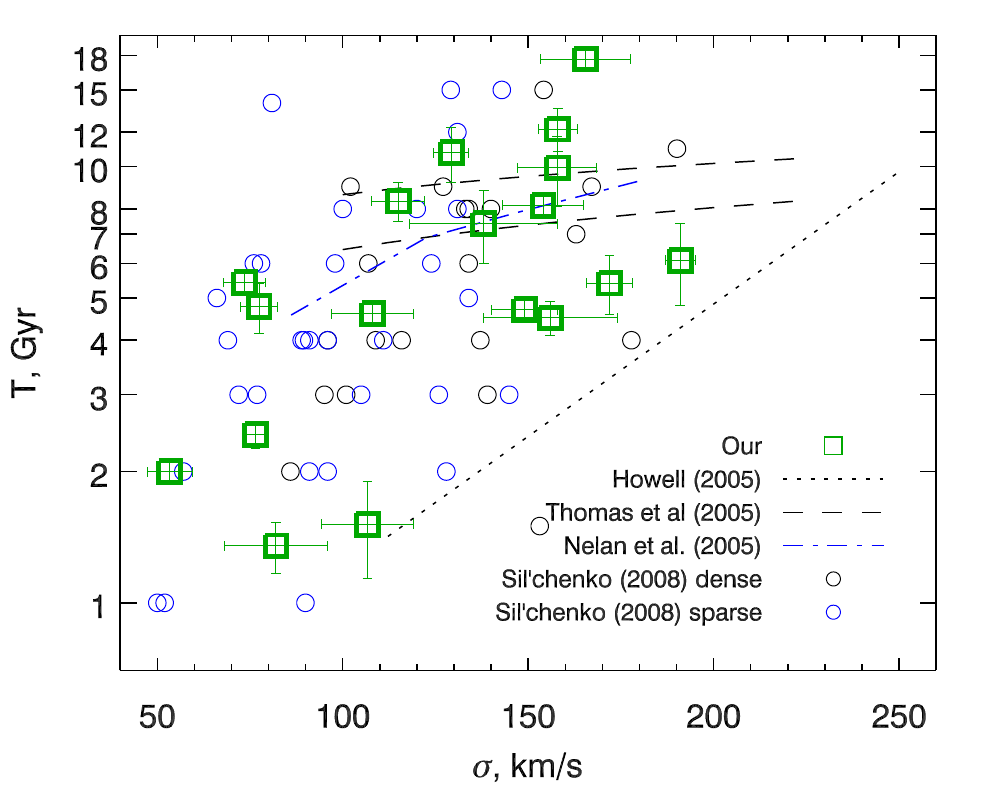}
\includegraphics[width=0.5\textwidth]{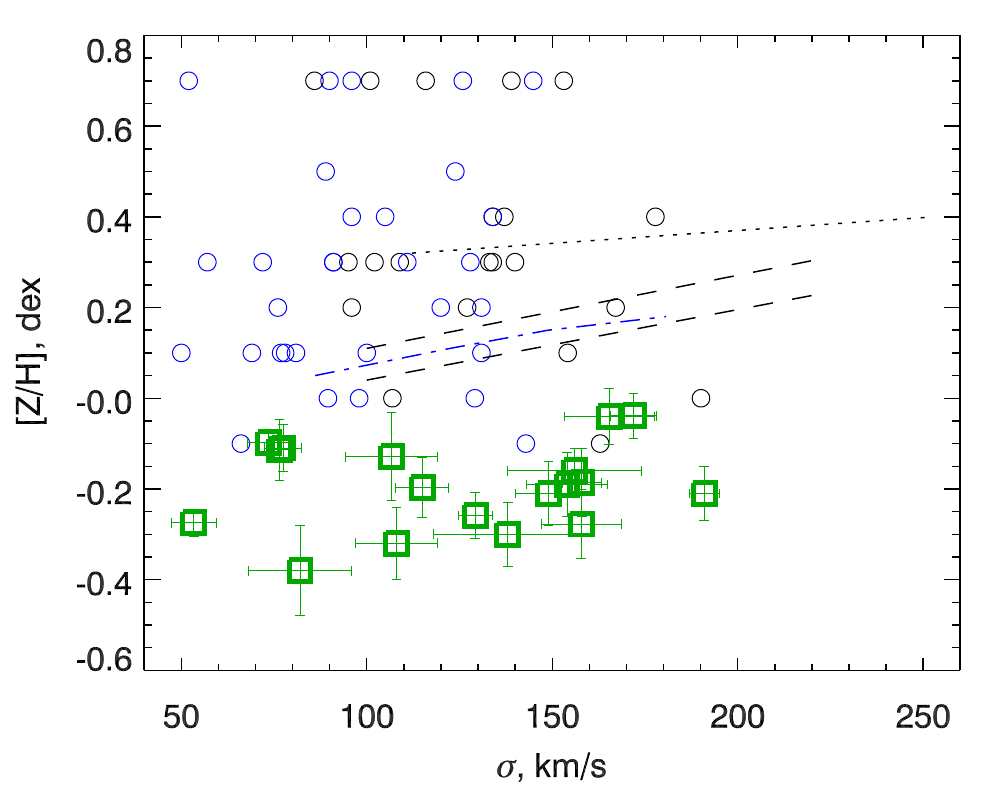}
}
\centerline{
\includegraphics[width=0.5\textwidth]{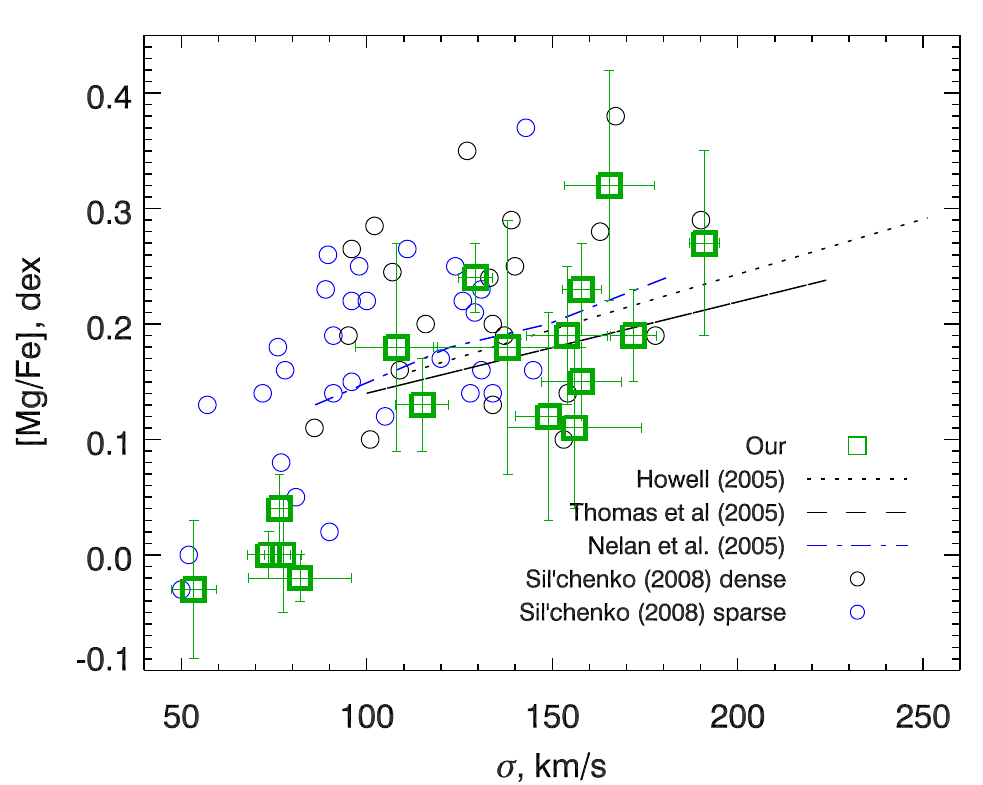}
}

\caption{Comparison of the stellar population properties with the stellar velocity dispersion
for the bulges of lenticular galaxies. Green squares shows our measurements in
the isolated lenticulars; circles correspond to measurements from
\citet{Silchenko2008_procIAU}, colors code environmental properties of
galaxies (\textit{black} - dense, \textit{blue} - sparse). Various samples of
early-type galaxies from literature are also shown. Details see in
\Se~\ref{stpop_bulge_comparison}.}\label{stpop_sig_comparison}

\end{figure*}

We can put our results on the stellar population properties in the bulges of 
isolated lenticular galaxies into a wider context by referring to the study of nearby
lenticular galaxies with the integral-field spectrograph of the Russian 6-meter telescope,
MPFS, by \citet{sil2006,Silchenko2008_procIAU}. In this work \citep{Silchenko2008_procIAU} the
data for the nuclei and bulges of more than 50 nearby S0s were presented; the sample
included galaxies over a wide range of environments and was divided into two parts --
`dense environments' (Virgo cluster and central galaxies of rich groups) and
`sparse environments' (mostly peripheries of groups). In Fig.~\ref{stpop_sig_comparison}
we reproduce the Fig.~2 from \citet{Silchenko2008_procIAU} where we overplot our present
results on the bulges of completely isolated S0s. Besides the data on the bulges of
nearby S0s, this figure contains also mean relations for the integrated stellar 
populations properties of elliptical galaxies in clusters \citep{nelan05}, in the
field \citep{howell}, and in both types of environments \citep{thomas05}. We must note
here that among these comparison samples only ellipticals were considered by \citet{howell};
\citet{nelan05} and \citet{thomas05} have mixed ellipticals and lenticulars. In
Fig.~\ref{stpop_sig_comparison} we see that the relations connecting the ages and the
[Mg/Fe] ratios with the stellar velocity dispersion found for spheroids in a wide
range of environments are broadly consistent with our results on the bulges of isolated
lenticular galaxies. However the stellar metallicities of the bulges are on average twice lower
in the isolated S0s with respect to all other samples. We can speculate that this
difference may be related to the possible difference in gas accretion sources in S0s
galaxies in different environments, if the SSP-equivalent metallicity of the bulges
is biased toward the metallicity of the last stellar generation born during some bulge
rejuvenation event.
\label{stpop_bulge_comparison}

% Figure 10 %%%%%%%%%%%%%%%%%%%%%%%%%%%%%%%%%%%%%%%%%%%%%%%%%%%%%%%%%%%%%%%%%%%%%%%%%%%%%%%%%%%%%%%%%%%

\subsection{Rings and lenses}

\begin{figure*}
\plottwo{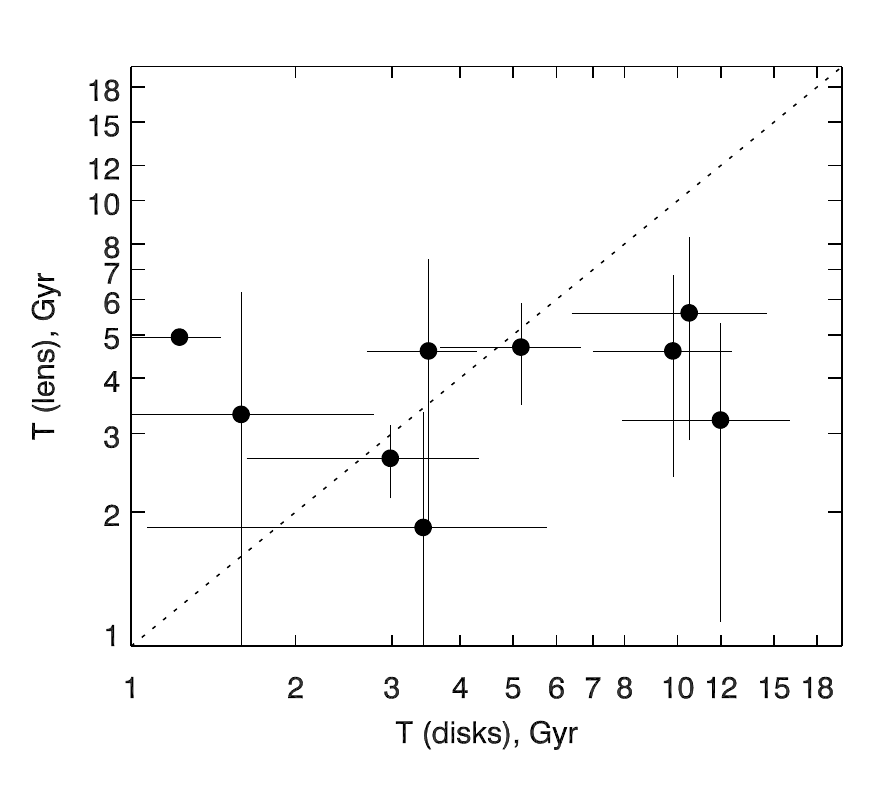}{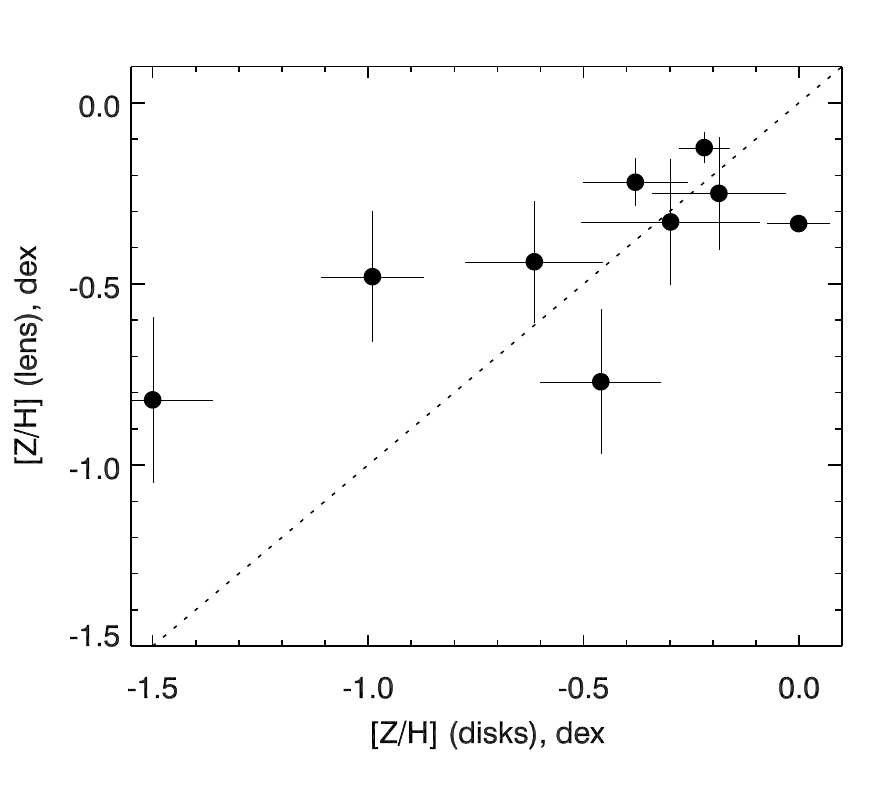}
\plottwo{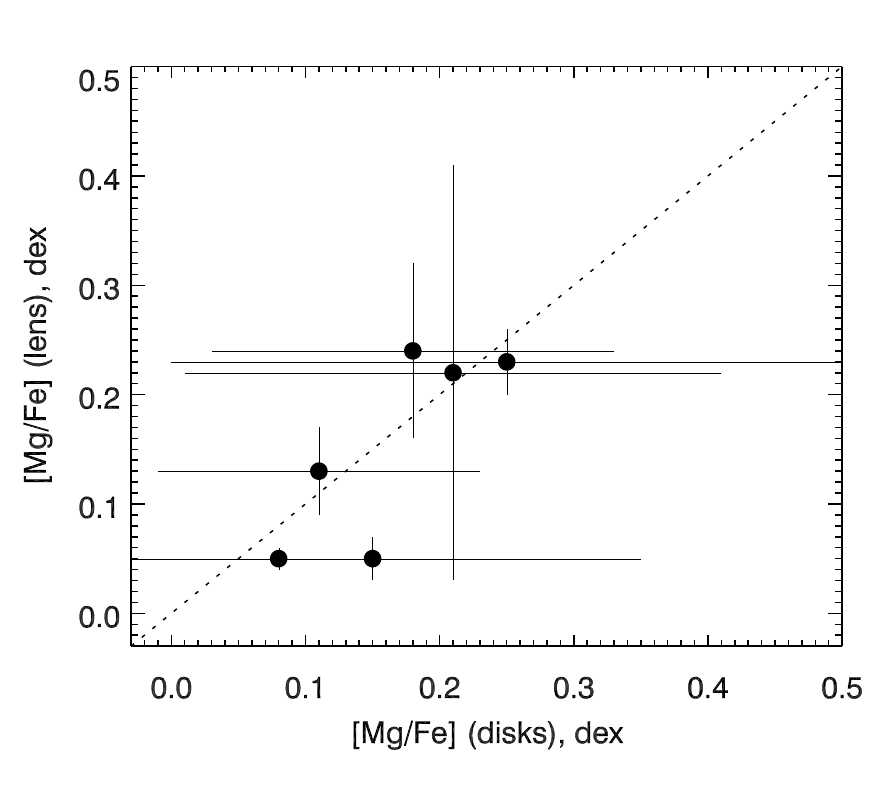}{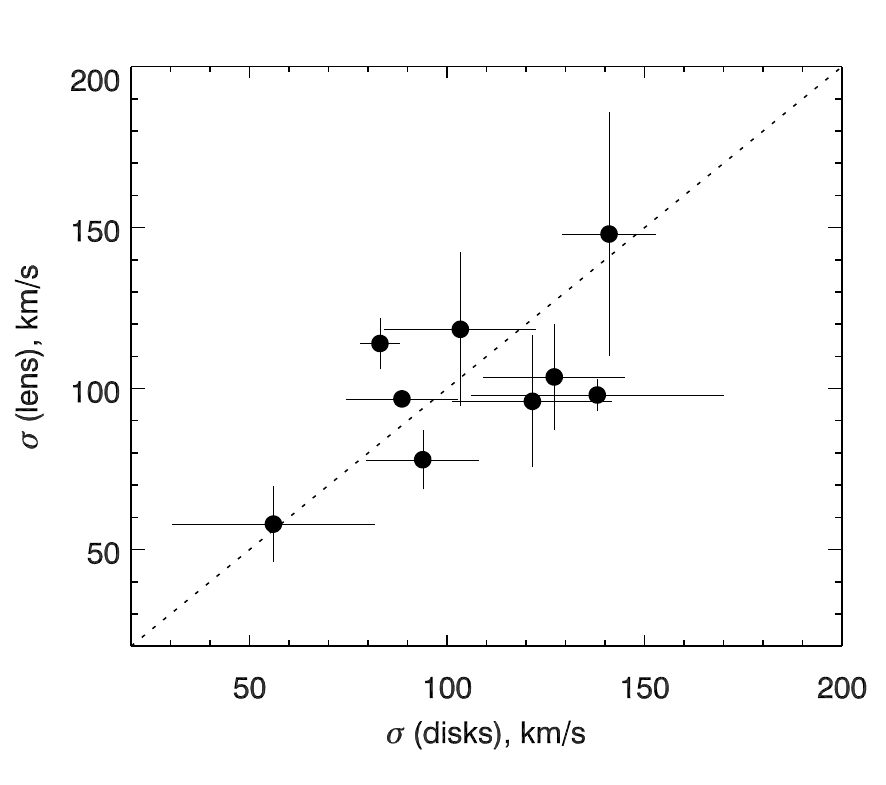}
\caption{Comparison of the disks and their rings/lenses. Dotted line
corresponds to equality line.}
\label{stpop_discs_lens}
\end{figure*}

General structure of lenticular galaxies differs from that of other disk
galaxies by often revealing such disk features as stellar rings (most frequent
in S0/a, \citet{deLapparent_2011}) and lenses (most frequent in S0s,
\citet{Laurikainen_2009}). It is a common view that the stellar lenses are
very old and dynamically hot though this point of view is based on very rare
observations of a few objects \citep{Kormendy_1984,Laurikainen_2013}.  We have
succeeded to measure kinematical and stellar population characteristics for 9
lenses and rings, and our results contradict to this common view. The stellar
velocity dispersions are generally the same in the lenses/rings and in the
surrounding disks (Fig.~\ref{stpop_discs_lens}, right bottom plot) so
dynamically they are indistinguishable; perhaps, there are some hints that the
rings and lenses can be found mostly in dynamically hot disks.
Fig.~\ref{stpop_discs_lens} gives also evidence for identical chemical
properties of the lenses/rings and their surrounding disks. But there is
however one important distinction between the disks and the lenses/rings:
while the mean stellar ages of the disks fill out the complete range of
possible values, between 1 and 12~Gyr, the ages of the rings and lenses are
predominantly concentrated in the narrow range between 2 and 6~Gyr
(Fig.~\ref{stpop_discs_lens}, left upper plot). An exception is the galaxy
from SCORPIO sample - NGC~6615 which has largest ring among entire sample with
the age of $\approx13$~Gyr. We can so state that the last starforming episodes
took place in these substructures at $z<1$. Here we see an association with
the fact that strong bars are predicted to appear in galactic disks only after
$z=1$ \citep{Kraljic_2012}. Since lenses in S0s are commonly related with
dissolved or weakened bars \citep{Buta_2010} and since star formation in rings
is usual at the resonance radii of the bars \citep{Buta_Comb_1996}, we would
like to connect the epoch of the last starforming episodes in the rings and
lenses and the epoch of the rapid bar arising in the stellar disks after
$z<1$.

% Figure 8 %%%%%%%%%%%%%%%%%%%%%%%%%%%%%%%%%%%%%%%%%%%%%%%%%%%%%%%%%%%%%%%%%%%%%%%%%%%%%%%%%%%%%%%%%%%

\begin{figure*}
\centerline{
    \includegraphics[width=0.33\textwidth]{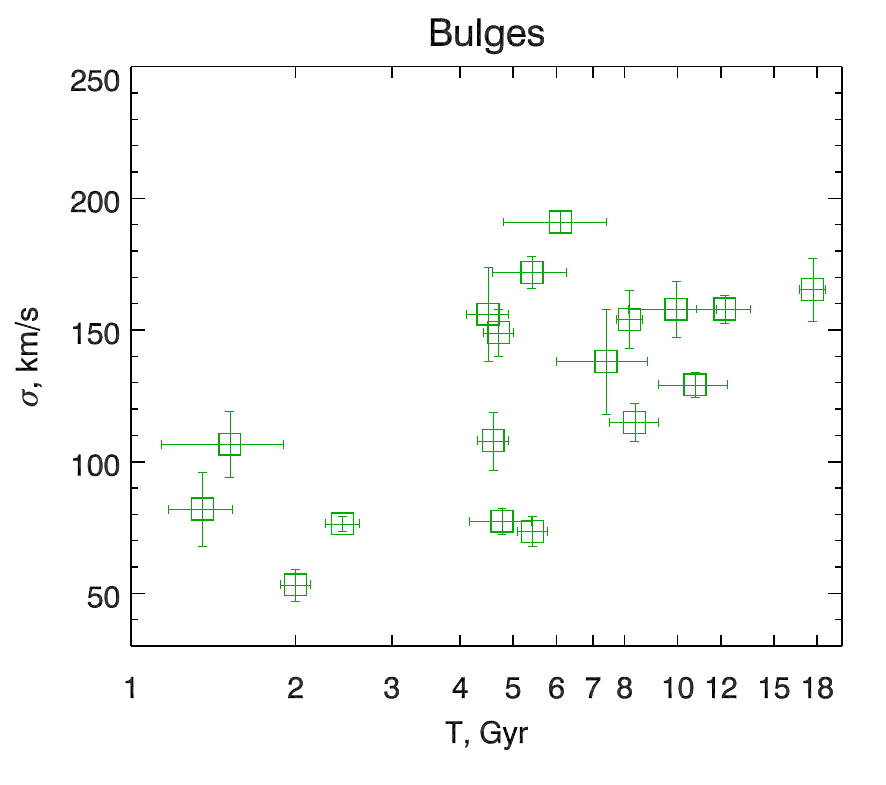}
    \includegraphics[width=0.33\textwidth]{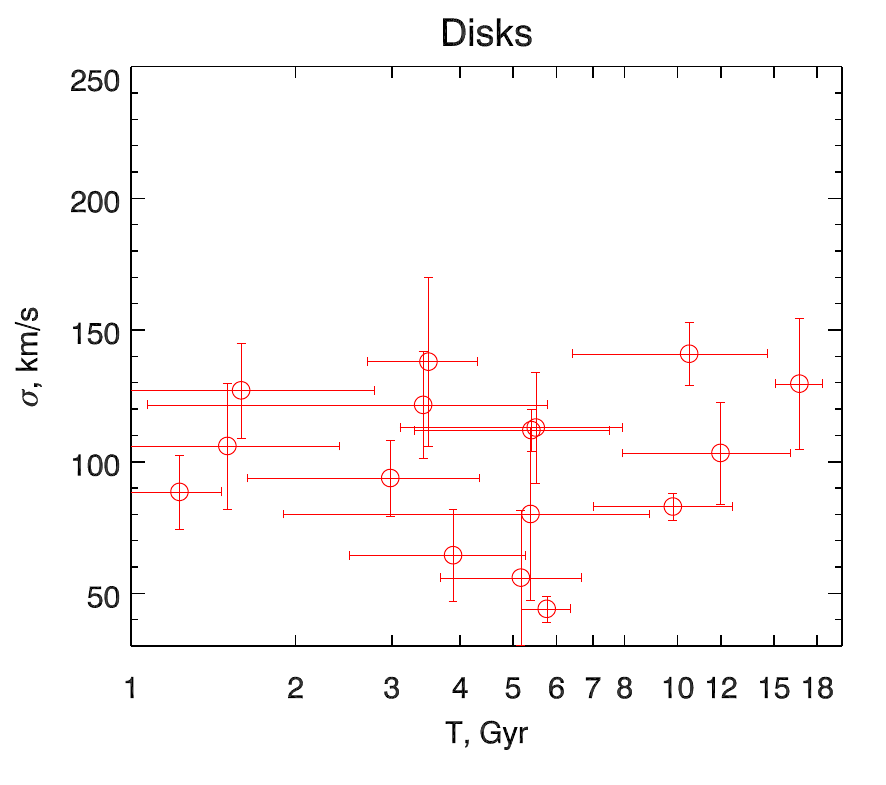}
    \includegraphics[width=0.33\textwidth]{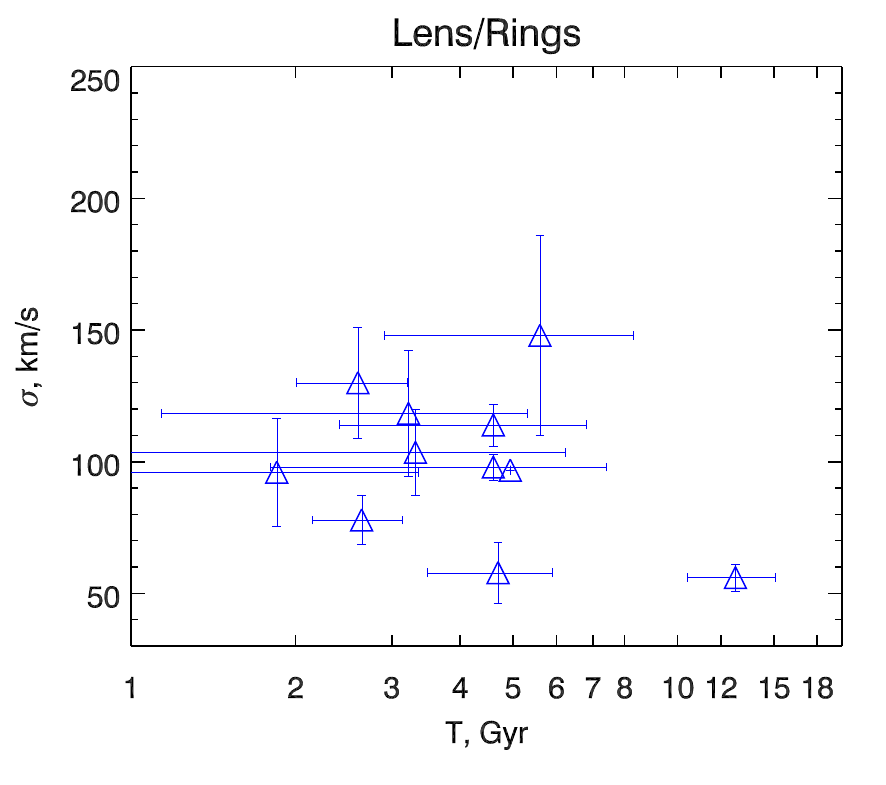}
}
\centerline{
    \includegraphics[width=0.33\textwidth]{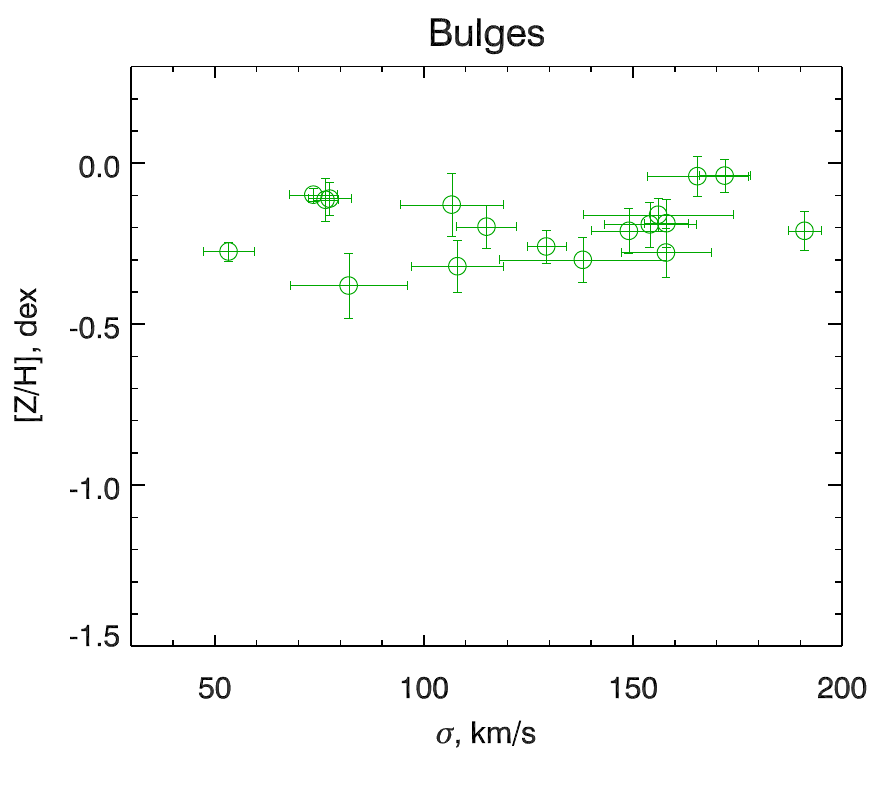}
    \includegraphics[width=0.33\textwidth]{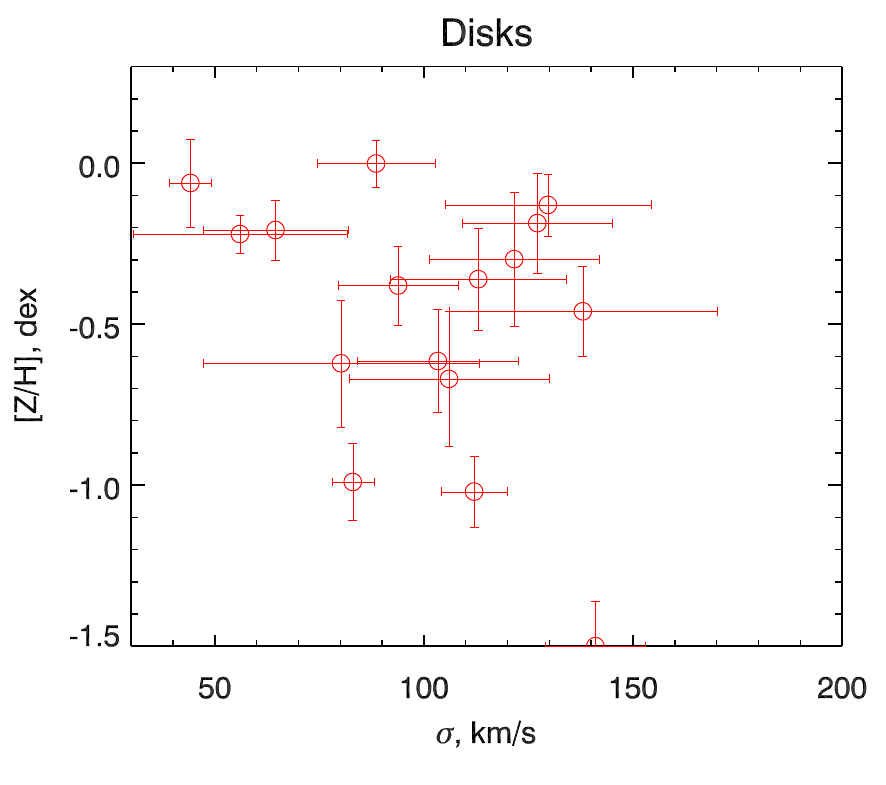}
    \includegraphics[width=0.33\textwidth]{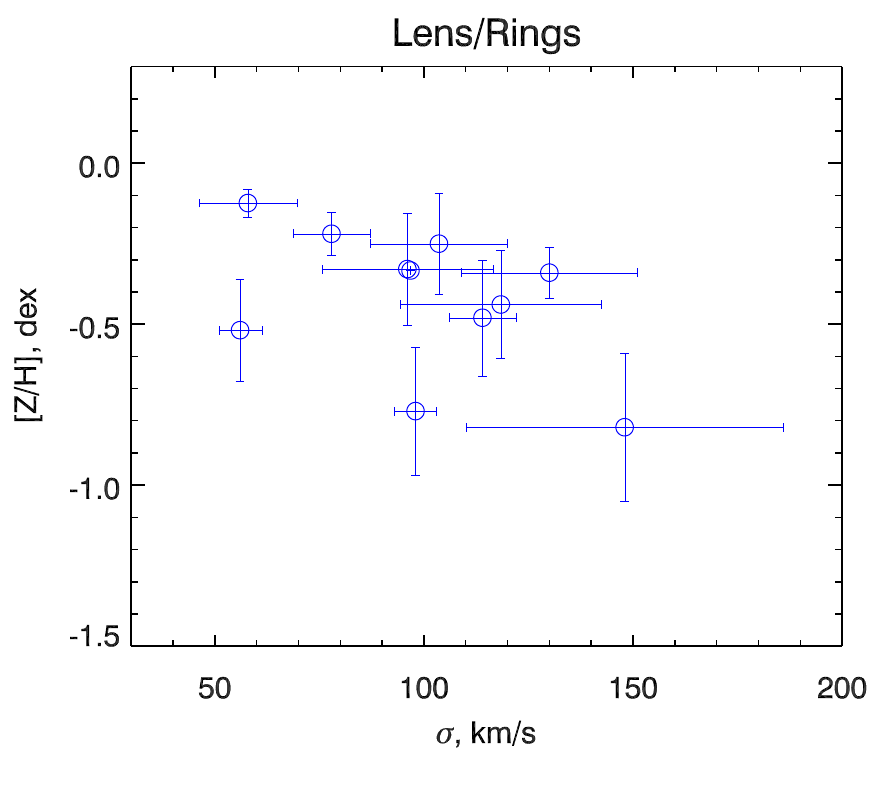}
}
\caption{Parameters of the stellar populations for the joined sample SALT \& SCORPIO targets.}
\label{stpopcomp}
\end{figure*}

Another interesting finding can be seen in Fig.~\ref{stpopcomp}. While the
bulges show the correlation between their ages and the stellar
velocity dispersion, just as is known for elliptical galaxies, the disks and
rings/lenses ages do not correlate with the observed stellar velocity
dispersion. These findings are supported by the evaluation of Spearman correlation 
coefficient between the ages and velocity dispersion of the bulges, $r_s=0.58$, with the
probability of the correlation to be insignificant $p=0.012$, while the correlation 
coefficients for the disks, $r_s=0.07$, $p=0.80$, and for the rings/lenses, $r_s=-0.17$, 
$p=0.61$, prove that here the dependencies are absent.
In addition, we found slight anticorrelation between the stellar metallicity and
velocity dispersion in the ring/lens structures ($r_s=-0.46$, $p=0.15$) while
for the disks and bulges such correlation is insignificant ($r_s=-0.26$, $p=0.34$
for the disks and $r_s=0.12$, $p=0.62$ for the bulges, correspondingly). It is obvious
that such anticorrelation, if exists, has an evolutionary census; but extension of the 
galaxy sample with reliable measurements of the stellar population properties in the
disk substructures is needed to strengthen the relation and to propose a
particular scenario to explain it.

% Figure 9 %%%%%%%%%%%%%%%%%%%%%%%%%%%%%%%%%%%%%%%%%%%%%%%%%%%%%%%%%%%%%%%%%%%%%%%%%%%%%%%%%%%%%%%%%%%

\subsection{Ionized-gas characteristics}

By analyzing the half of our sample observed at the Russian 6m telescope, we
have noted that, firstly, the majority of isolated lenticular galaxies contain
extended ionized-gas disks, and secondly, the rotation and orientation of the
ionized-gas disks are often decoupled from the rotation and orientation of the
stellar disks \citep{ilg_gas}. Now, with the complete sample in hands, we can
refine the statistics of the ionized-gas content of the isolated lenticular
galaxies. Among 18 galaxies studied, 13 galaxies demonstrate extended
ionized-gas emission (72\%$\pm 11$\%); and among 13 galaxies with the extended
gas emission, 7 galaxies (54\%$\pm 14$\%) demonstrate visible counterrotation
of the ionized gas with respect to their stellar components.  Our spectral
observations are `one-dimensional': the long slit aligned with the major axis
of the continuum isophotes characterizing the line of nodes of the {\it
stellar} disk cannot help to determine the orientation of the {\it gas}
rotation plane. Following the logic proposed by \citet{bertola92}, if we
suppose that the gas in S0s is accreted from external sources, and the orbital
momentum of the accreted gas is oriented accidentally with respect to the
angular momentum of the galaxy, we should see equal proportions of corotating
and counterrotating gas by studying only the gas velocity projection onto the
stellar disk lines of nodes.  It is just what we have found from our
observations of the isolated lenticular galaxies. So we may conclude that the
statistics of the ionized-gas rotation in our sample of the isolated lenticular
galaxies gives evidence for the {\it all} gas having been accreted from
external sources isotropically distributed around the galaxies.

What can these sources be? Since our galaxies are {\it isolated} and do not
have neighbouring large galaxies which may be donors of the gas, we can propose
only two probable sources of the decoupled gas acquisition: minor merging of
small gas-rich satellites
\citep{Reshetnikov_Sotnikova_1997,Bournaud_Combes_2003} or gas inflow from
cosmological filaments of the Universe large-scale structure
\citep{keres_flows,dekel_flows,Bournaud_Elmegreen_2009}.  We suggest that
metallicity of the gas can help to identify exactly the gas origin:
cosmological filaments of the Universe large-scale structure must contain the
pristine gas so it must be very metal-poor \citep{Agertz_2009}. By pursuing
this aim, we have picked out in our galaxies the radial ranges along the slit
where the ionized gas is excited by young stars, according to the BPT-diagram
diagnostics, and then we have added the spectra over these ranges for every
galaxy. To these rather high-S/N spectra, we have applied so called
`strong-line calibrations' allowing to estimate oxygen abundance in the
HII-regions with only a few emission lines, namely, with the Balmer lines
H$\alpha$ and H$\beta$, low-excitation [NII]$\lambda$6583, and high-excitation
[OIII]$\lambda$5007.  We have succeeded to estimate metallicities of the
ionized gas in 8 isolated lenticular galaxies. The results are presented in the
Table~\ref{table_abund}. Despite the wide range of galaxy luminosities, the
ionized-gas metallicities have appeared to be confined to a very narrow range
of values near the solar metallicity or slightly higher. So we think that we
can exclude cosmological filaments as the source of gas accretion in this
particular case. Obviously, we see the consequences of gas-rich satellite
merging.

% Table 4 %%%%%%%%%%%%%%%%%%%%%%%%%%%%%%%%%%%%%%%%%%%%%%%%%%%%%%%%%%%%%%%%%%%%%%%%%%%%%%%%%%%%

\begin{deluxetable}{rcc}
\tablecolumns{3}
\tablewidth{0pc}

\tablecaption{Estimates of the ionized-gas oxygen abundance in the emission regions
excited by young stars.  $N2$ marks the estimations which has been done by
using only $N2$ emission line index.\label{table_abund}}

\tablehead{
\colhead{Galaxy} & \colhead{Radial range of}  & \colhead{12+$\log$ O/H ([Z/H]$_O$),} \\
                 & \colhead{binning, arcsec}  & \colhead{dex}                    }

\tabletypesize{\scriptsize}

\startdata
\multicolumn{3}{c}{SALT-data}\\
\hline
\multirow{2}{*}{   IC 1608} & (    -51.4;     -31.1) &       8.78 (   0.09)  $\pm   0.47$ \\
                              & (     29.7;      43.4) &       8.80 (   0.11)  $\pm   0.26$ \\
\hline
%\multirow{2}{*}{   IC 3152$^{N2}$} & (    -12.6;      -8.1) &       8.73 (   0.04)  $\pm   0.42$ \\
%                              & (      4.1;      10.2) &       8.78 (   0.09)  $\pm   0.42$ \\
%\hline
\multirow{2}{*}{  NGC 1211$^{N2}$} & (      9.4;      15.8) &       8.72 (   0.03)  $\pm   0.41$ \\
                              & (     32.3;      37.3) &       8.73 (   0.04)  $\pm   0.41$ \\
\hline
\multirow{2}{*}{  NGC 2917} & (    -15.7;      -5.6) &       8.90 (   0.21)  $\pm   0.27$ \\
                              & (      9.6;      19.8) &       8.82 (   0.13)  $\pm   0.26$ \\
\hline
\multirow{2}{*}{  NGC 4240$^{N2}$} & (    -11.6;      -6.6) &       8.80 (   0.11)  $\pm   0.41$ \\
                              & (      4.1;      12.2) &       8.78 (   0.09)  $\pm   0.41$ \\
\hline
\multirow{2}{*}{  UGC 9980$^{N2}$} & (    -13.1;      -4.3) &       8.82 (   0.13)  $\pm   0.42$ \\
                              & (      8.4;      20.8) &       8.71 (   0.02)  $\pm   0.42$ \\
\hline
\multicolumn{3}{c}{SCORPIO-data}\\
\hline
  NGC 2350 & (     -1.6;       2.0) &       8.68 (  -0.01)  $\pm   0.25$ \\
\hline
\multirow{2}{*}{  NGC 6798$^{N2}$} & (    -34.1;     -27.7) &       8.71 (   0.02)  $\pm   0.41$ \\
                              & (     29.1;      36.6) &       8.73 (   0.04)  $\pm   0.41$ \\
\hline
  NGC 7351 & (     -2.7;       3.8) &       8.64 (  -0.05)  $\pm   0.25$ \\

\enddata

\end{deluxetable}

Though formally we cannot determine the orientation of the ionized-gas rotation
plane with the only long-slit spectroscopy, we can note some possible signatures of the
gas confinement to the plane of the stellar disk: it may be consistency of the
rotation velocity estimates for the stars and for the gas in the outer parts of the 
galaxies where the stellar velocity
dispersion is low and does not affect strongly the line-of-sight velocity
profiles through the asymmetric drift.  Among our sample, such consistency is
demonstrated by the galaxies with co-rotating ionized gas IC 1608, NGC 1211,
NGC 2350, NGC 2917, UGC 9980, and by the galaxies with the {\it counterrotating}
ionized gas NGC 4240 and NGC 6798; just these galaxies figure in the
Table~\ref{table_abund} with their outer ionized gas excited by young stars. Other
galaxies where we can suspect the gas rotating off the main symmetry plane show
mostly other types of excitation -- by shock waves or by old post-AGB stars placing
the emission-line flux ratios at the BPT-diagrams to the right from the
dividing curve. We can here remind theoretical consideration by
\citet{wakamatsu93} who noted that inclined gaseous disks/rings must experience
shocks developed because of the gas crossing gravitational potential well of a
stellar disk. The shock waves must heat the gas and prevent its cooling
necessary for a star formation burst. We suggest then that conditions for star
formation starting in the accreted gas must include, besides the gas amount,
also the favorable geometry of the gas accretion.

\subsection{Origin of isolated S0s}

Just from this point we would like to start discussion about the origin of
isolated lenticular galaxies.  Indeed, all the mechanisms proposed so far to
quench star formation in the disks of spiral galaxies and to transform them
into lenticulars act only in dense environments -- in clusters and massive rich
groups.  It remained so quite unclear how lenticular galaxies in the field
might form. The situation changes if we accept the new paradigm for evolution
of disk galaxies proposed by \citet{sil_s0}: all disk galaxies started their
evolution as lenticulars at the redshifts of $z=2-3$, and only after $z<1$ most
of them became spirals by undergoing persistent outer gas accretion onto their
disks that resulted in dynamical cooling and subsequent spiral-arm development
and star formation ignition. Then the key point for galaxy morphological
shaping becomes outer-gas accretion regime. In clusters the outer cold gas
accretion is almost impossible due to tide-induced starvation and hot
intracluster medium ram pressure so in clusters the most disk galaxies remain
lenticulars for all their lifes. In the field the conditions for outer cold gas
accretion can be quite various. If we regard gas-rich satellite merging as a
main outer gas source, then we can expect the following variety of satellite
system properties for the isolated disk galaxies: How many satellites has the
host galaxy? Are they distributed isotropically around it or are confined to
some dedicated plane as it is observed in our Galaxy and in M~31? Is the
satellite system dynamically cold or hot (related perhaps to the mass of the
host dark matter halo)?  Concerning the last point, there is a curious fact
noticed by \citet{kara11}. They considered faint companions of isolated
galaxies from their catalog 2MIG. They found that the companions of isolated
{\it early-type} galaxies have in average larger line-of-sight velocity
difference with their hosts than the late-type ones. By inspecting their Fig.~4
we have ascertained that there is no practically companions of isolated
early-type galaxies with the line-of-sight velocity difference less than 50
\kms\ while the isolated late-type hosts possess a lot of such companions. It
is a natural suggestion that the accretion of companions with a large flyby
velocity is more difficult than that of slow ones, so we come to a conclusion
that perhaps just a `hot' orbital population of companions defines an early
morphological type of the host isolated galaxy. Perhaps, the orbital
composition of a satellite system relates stochastically to initial conditions,
or may be a present-day isolated lenticular have merged all its slow companions
several Gyr ago and has no ones presently... In other cases, accretion of a
gas-rich satellite from a highly inclined orbit may lead to gas heating and
prevent star formation and spiral-arm development which requires cold,
gravitationally unstable disk. By varying possible variants of satellite
merging regime, we can easily get an isolated galaxy of any morphological type
-- in opposition to the tight accretion conditions in dense environments
which provide a subsequent tight range of morphological types, mostly S0s.

\section{Summary}\label{txt:summ}

We have observed 9 galaxies from our sample of isolated lenticular galaxies at
the 10m Southern African Large Telescope with the Robert Stobie Spectrograph in
the long-slit mode. The radial variations of the kinematical characteristics,
line-of-sight velocities and velocity dispersions, are studied both for the
stellar component and for the ionized gas which is found in the most part of
the sample. Also we have derived radial profiles of the mean stellar
metallicity and ages, as well as the gas excitation characteristics and oxygen
abundances far outward, into the disk-dominated regions of the galaxies.  By
joining two subsamples of isolated lenticular galaxies studied by us here and
earlier, the northern and southern ones, with a totality of 18 isolated
lenticular galaxies, we analyze the statistics of the stellar population
properties and ionized-gas features for this morphological type of galaxies in
extremely rarefied environments.

We have found that there is no particular time frame for shaping the isolated
lenticular galaxies: the mean stellar ages of the bulges and disks are
homogeneously distributed between 1 and $>13$~Gyr, and the bulges and disks
tend to form synchronously having mostly similar ages and magnesium-to-iron
ratios. In some galaxies we have found stellar disk substructures -- rings and
lenses; their mean stellar ages are confined to a rather narrow range, from 2
to 5 Gyr. We relate the appearance of these structures to strong bars arising
in disk galaxies after $z<1$.

An ionized-gas extended emission is found in the majority of our galaxies, in
13 of 18 (72\%$\pm 11$\%).  And the half of all extended gaseous disks
demonstrate visible counterrotation with respect to their stellar counterparts.
Just this proportion is expected if all the gas in isolated lenticular galaxies
is accreted from isotropically distributed external sources. A very narrow
range of the oxygen abundances, [O/H] from 0.0 to $+0.2$~dex estimated by us
for the outer ionized-gas disks excited by young stars, gives evidences for the
satellite merging as the most probable source of this accretion. At last we
formulate a hypothesis that morphological type of a field disk galaxy is
completely determined by the outer cold gas accretion regime.

\section*{Acknowledgments}

The observations reported in this paper were obtained with the Southern African
Large Telescope (SALT).  AYK acknowledges the support from the National
Research Foundation (NRF) of South Africa. The study of isolated lenticular
galaxies is supported by the grant No.~13-02-00059a of the Russian Foundation for
Basic Research. IYK is grateful to Dmitry Zimin's non-profit Dynasty Foundation.

Funding for SDSS-III has been provided by the Alfred P. Sloan Foundation, the
Participating Institutions, the National Science Foundation, and the U.S.
Department of Energy Office of Science. The SDSS-III web site is
http://www.sdss3.org/. SDSS-III is managed by the Astrophysical Research
Consortium for the Participating Institutions of the SDSS-III Collaboration
including the University of Arizona, the Brazilian Participation Group,
Brookhaven National Laboratory, Carnegie Mellon University, University of
Florida, the French Participation Group, the German Participation Group,
Harvard University, the Instituto de Astrofisica de Canarias, the Michigan
State/Notre Dame/JINA Participation Group, Johns Hopkins University, Lawrence
Berkeley National Laboratory, Max Planck Institute for Astrophysics, Max
Planck Institute for Extraterrestrial Physics, New Mexico State University,
New York University, Ohio State University, Pennsylvania State University,
University of Portsmouth, Princeton University, the Spanish Participation
Group, University of Tokyo, University of Utah, Vanderbilt University,
University of Virginia, University of Washington, and Yale University.

%%%%%%%%%%%%%%%%%%%%%%%%%%%%%%%%%%%%%%%%%%%%%%%%%%%%%%%%%%%%%%%%%%%%%%%%%%%%%%%%%%%%%%%%%%%%%%%%%%%%%%%
% BIBLIOGRAPHY %%%%%%%%%%%%%%%%%%%%%%%%%%%%%%%%%%%%%%%%%%%%%%%%%%%%%%%%%%%%%%%%%%%%%%%%%%%%%%%%%%%%%%%%
\bibliographystyle{aj}
%\bibliography{ref_katkov}  
\bibliography{ilg_salt.bbl}  

\clearpage
\newpage
\begin{appendix}

\begin{figure}
\centerline{
	\includegraphics[width=\textwidth]{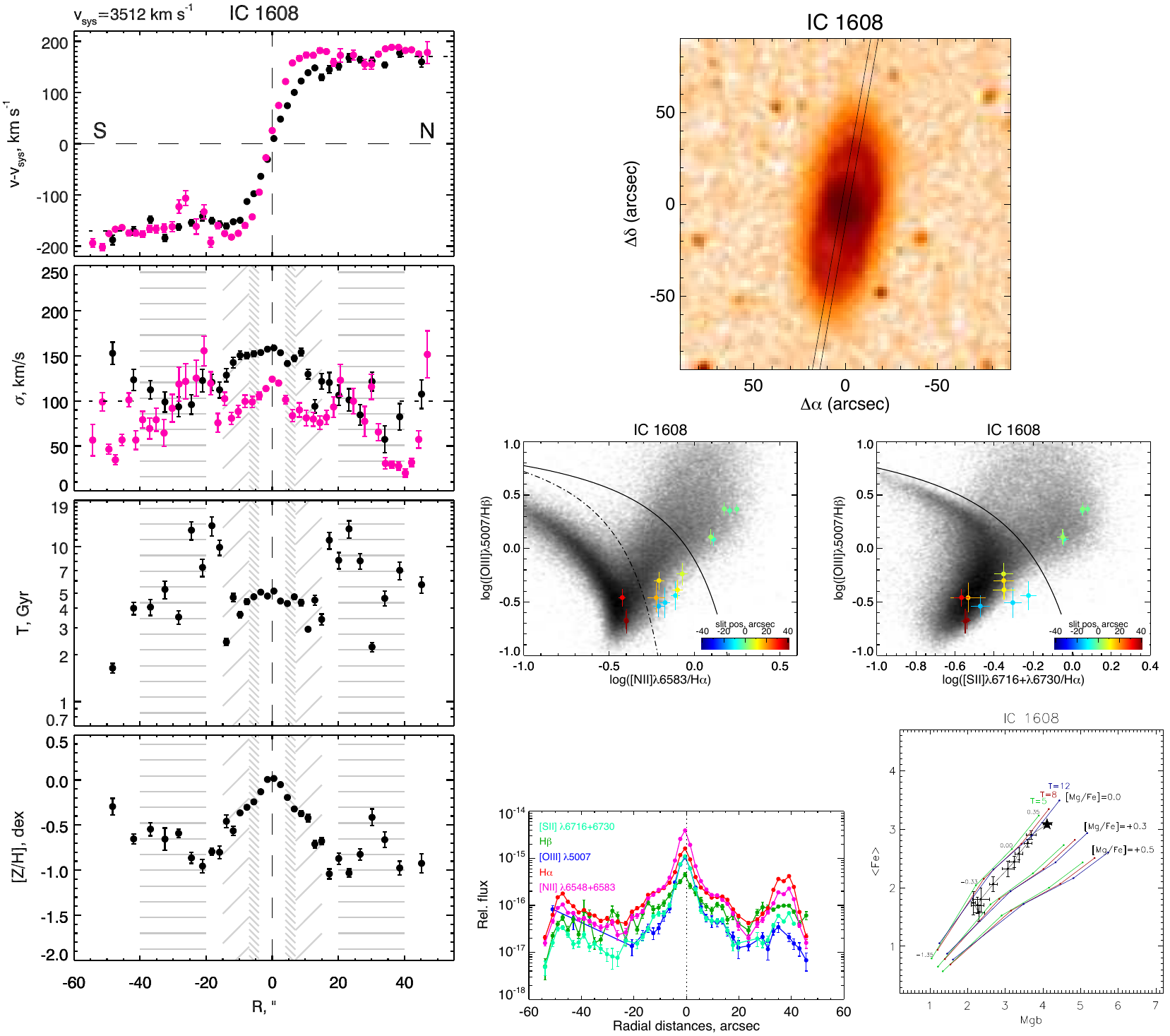}
}
\caption{\footnotesize{} IC 1608. \textit{Left block from top to bottom:}
\textit{i.} The radial profile of stellar (\textit{black}) and gaseous
(\textit{pink}) line-of-sight velocity; \textit{ii.} The stellar and gaseous
velocity dispersion; \textit{iii.} Properties of stellar populations -- ages
and metallicities. The shaded gray lines show radial segments where average
stellar population parameters are calculated: `\textbackslash'-like shading corresponds to
bulge dominance regions, `-' - to lens/ring regions, `/' - to disk ones.
\textit{Right block from top to bottom:} \textit{i.} Long-slit
position superimposed on an (DSS) image of galaxy; \textit{ii.} Excitation
diagnostic diagrams comparing different emission-line rations. The
reference distribution of the measurements of the line ratios for
galaxies from the SDSS survey with high signal-to-noise ratios (S/N $>$ 3 in
every line) are shown by gray color. The black curves, which separate the
areas with the AGN/LINER excitations from areas with the star-formation-
induced excitation, are taken from \citet{Kauffman_2003} (\textit{dash-dotted
curve}) and from \citet{Kewley_2006} (\textit{solid curve}). \textit{iii.}
Observed emission line fluxes. \textit{iv.} Diagnostic diagram $\left\langle
{Fe} \right\rangle$ versus Mgb. Points with error bars represent our measurements
along the radius of the galaxy, starting from the nucleus marked by a large star.
The SSP models by \citet{Thomasstpop} for three different magnesium-to-iron ratios 
(0.0, +0.3 and +0.5) and three different ages (5, 8 and 12 Gyr) are plotted 
as reference.} \end{figure} %%%%%
%%%%%%%%%%%%%%%%%%%%%%%%%%%%%%%%%%%%%%%%%%%%%%%%%%%%%%%%%%%%%%%%%%%%%%%%%%%%%%
%%%%%%%%%%%%%%%%%%%

\newpage
\begin{figure}
\centerline{
	\includegraphics[width=\textwidth]{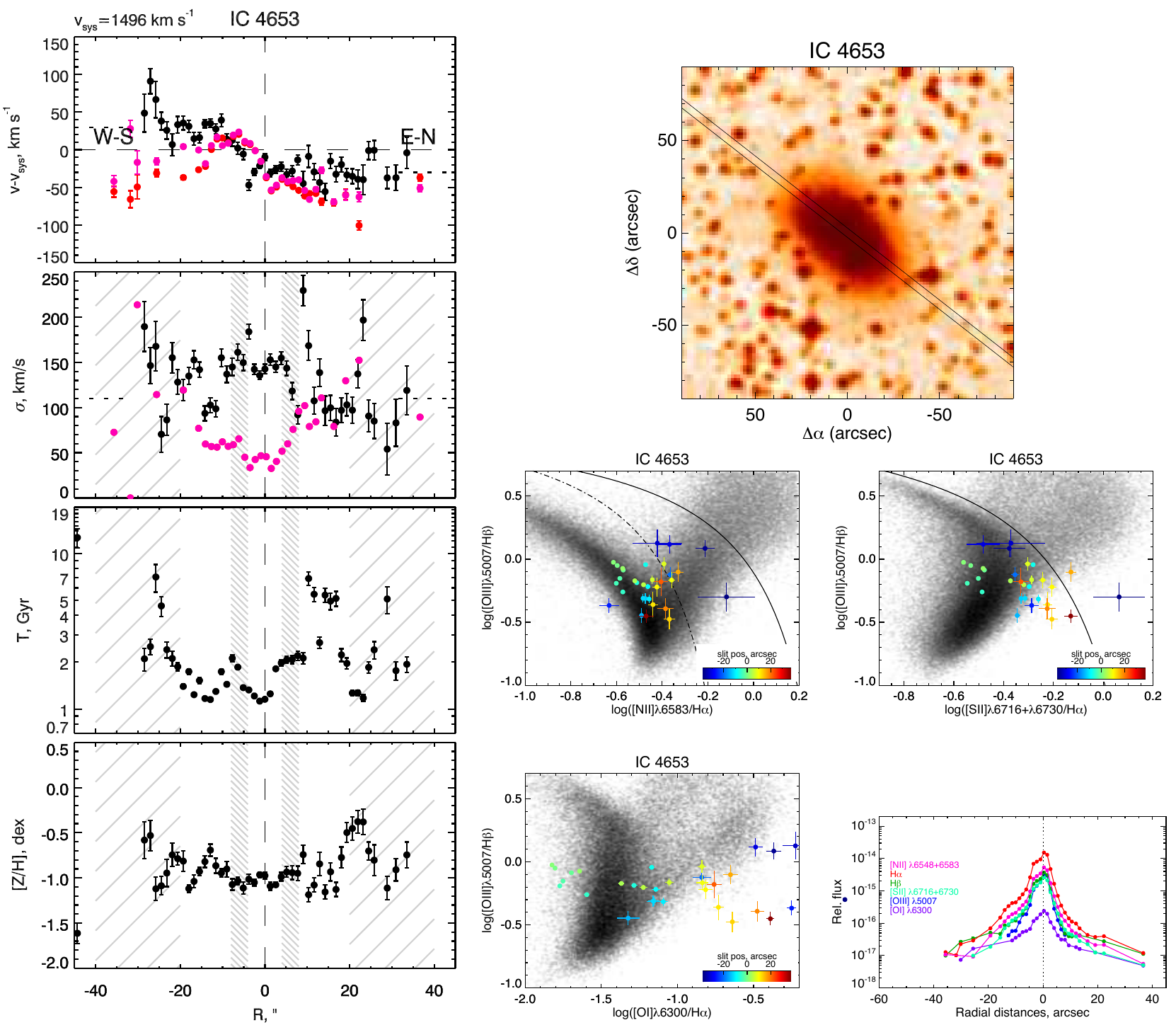}
}
\caption{IC~4653. The same as previous figure.}
\end{figure} 

% %%%%%%%%%%%%%%%%%%%%%%%%%%%%%%%%%%%%%%%%%%%%%%%%%%%%%%%%%%%%%%%%%%%%%%%%%%%%%%%%%%%%%%%%%%%%%%%%%%%%%%
\newpage
\begin{figure}
\centerline{
	\includegraphics[width=\textwidth]{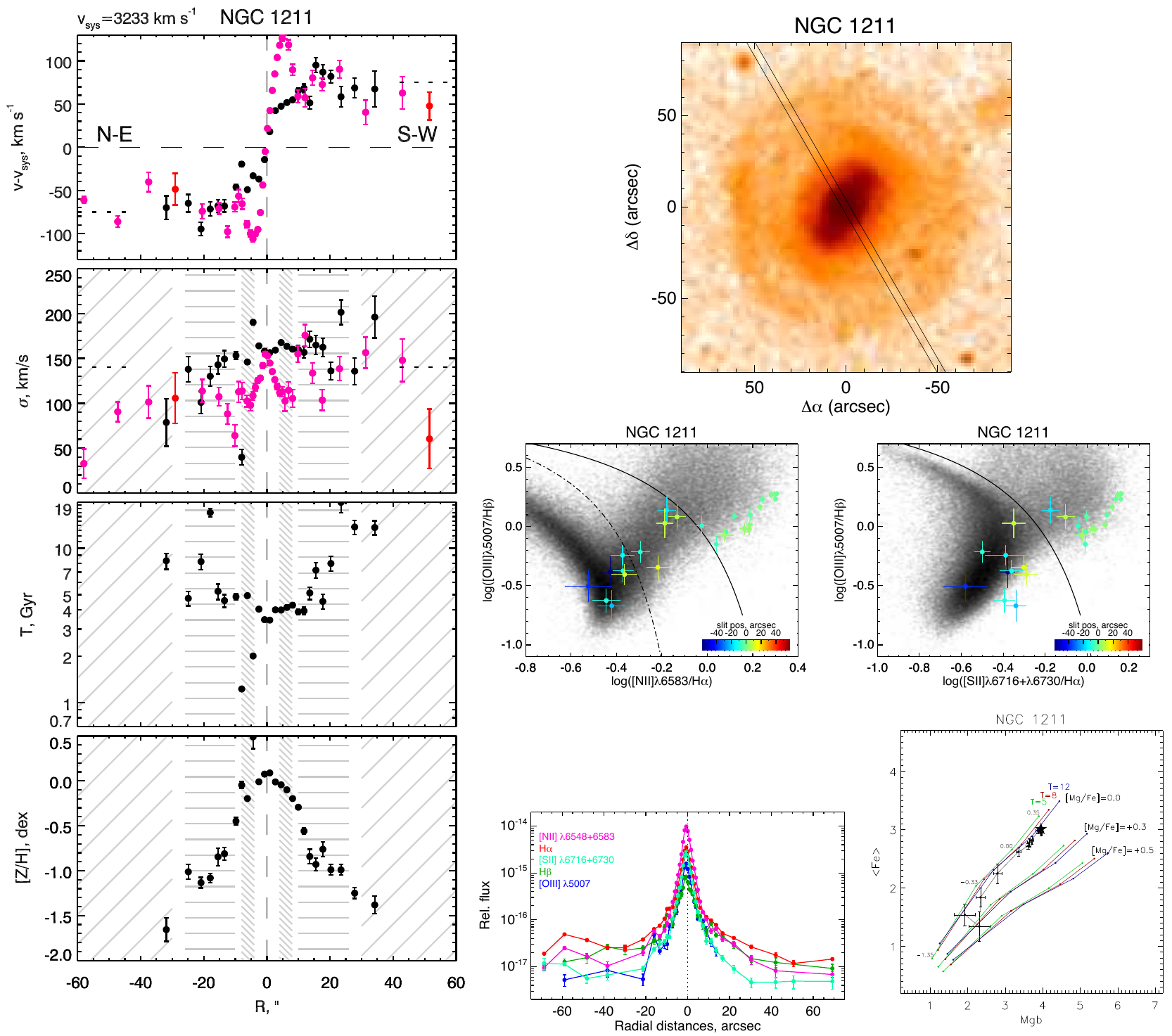}
}
\caption{NGC 1211. The same as previous figure.}
\end{figure}

%%%%%%%%%%%%%%%%%%%%%%%%%%%%%%%%%%%%%%%%%%%%%%%%%%%%%%%%%%%%%%%%%%%%%%%%%%%%%%%%%%%%%%%%%%%%%%%%%%%%%%
\newpage
\begin{figure}
\centerline{
	\includegraphics[width=\textwidth]{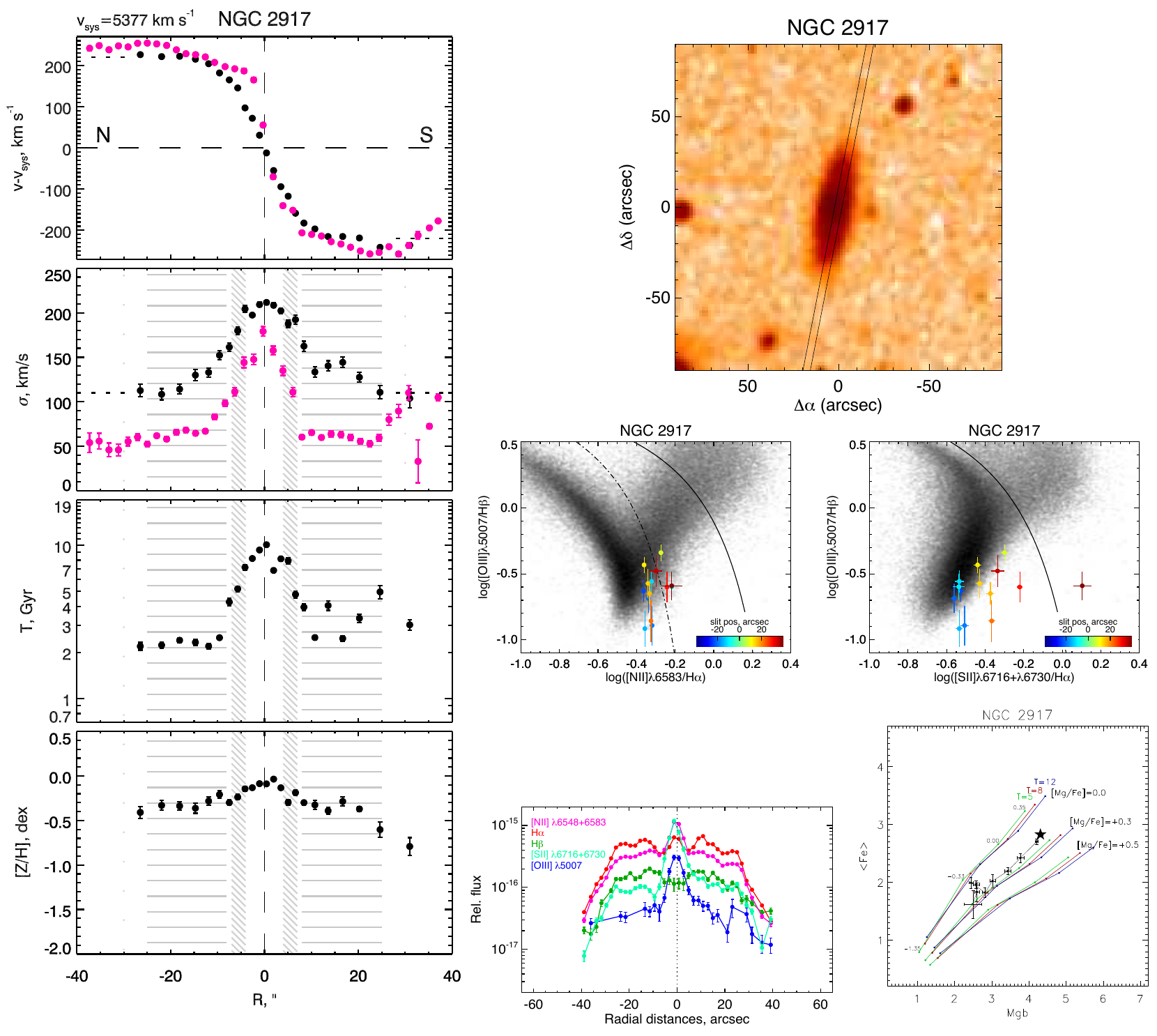}
}
\caption{NGC~2917. The same as previous figure.}
\end{figure} 

%%%%%%%%%%%%%%%%%%%%%%%%%%%%%%%%%%%%%%%%%%%%%%%%%%%%%%%%%%%%%%%%%%%%%%%%%%%%%%%%%%%%%%%%%%%%%%%%%%%%%%
\newpage
\begin{figure}
\centerline{
	\includegraphics[width=\textwidth]{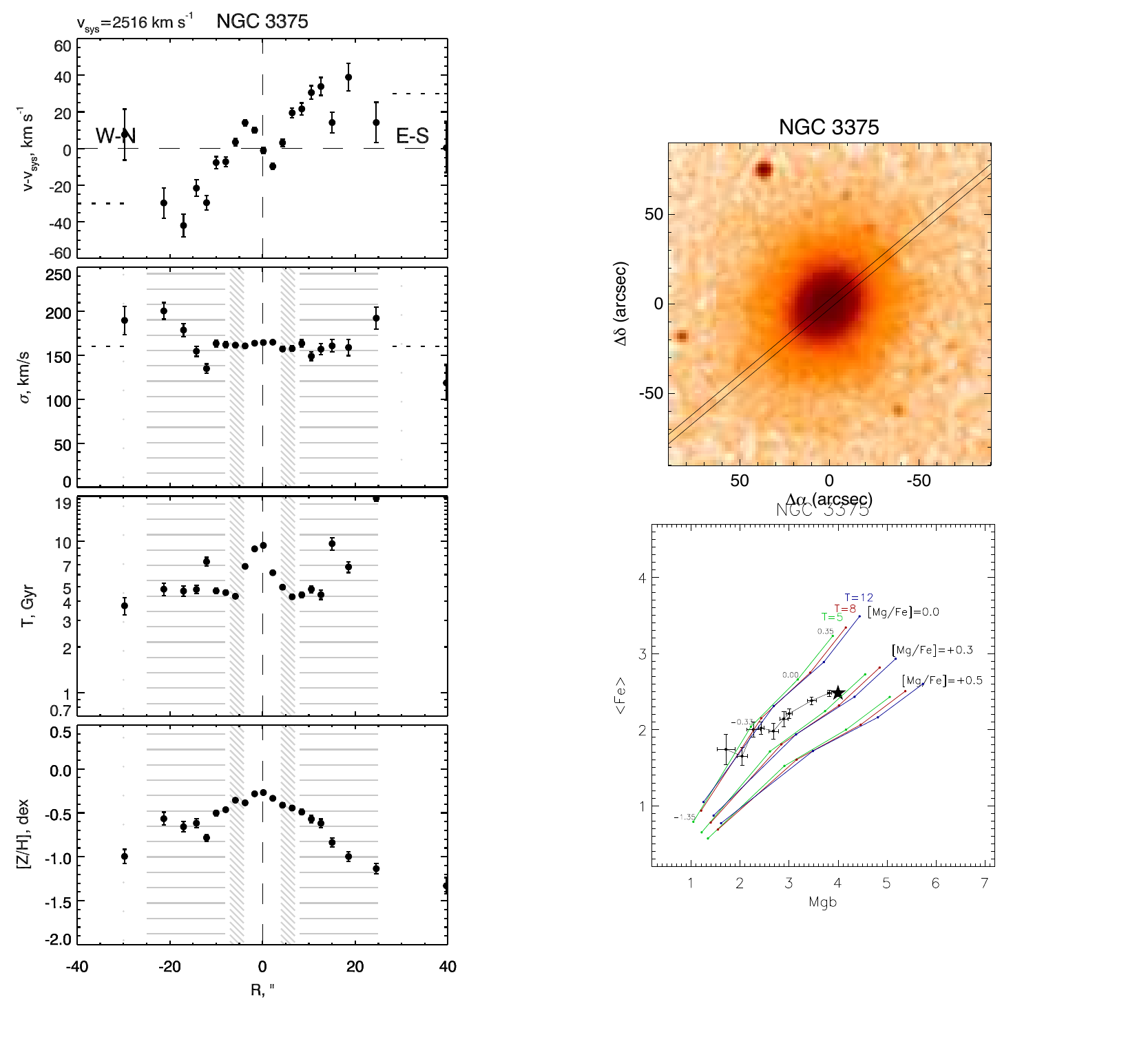}
}
\caption{NGC~3375. The same as previous figure.}
\label{fig_n3375}
\end{figure}

%%%%%%%%%%%%%%%%%%%%%%%%%%%%%%%%%%%%%%%%%%%%%%%%%%%%%%%%%%%%%%%%%%%%%%%%%%%%%%%%%%%%%%%%%%%%%%%%%%%%%%
\newpage
\begin{figure}
\centerline{
	\includegraphics[width=\textwidth]{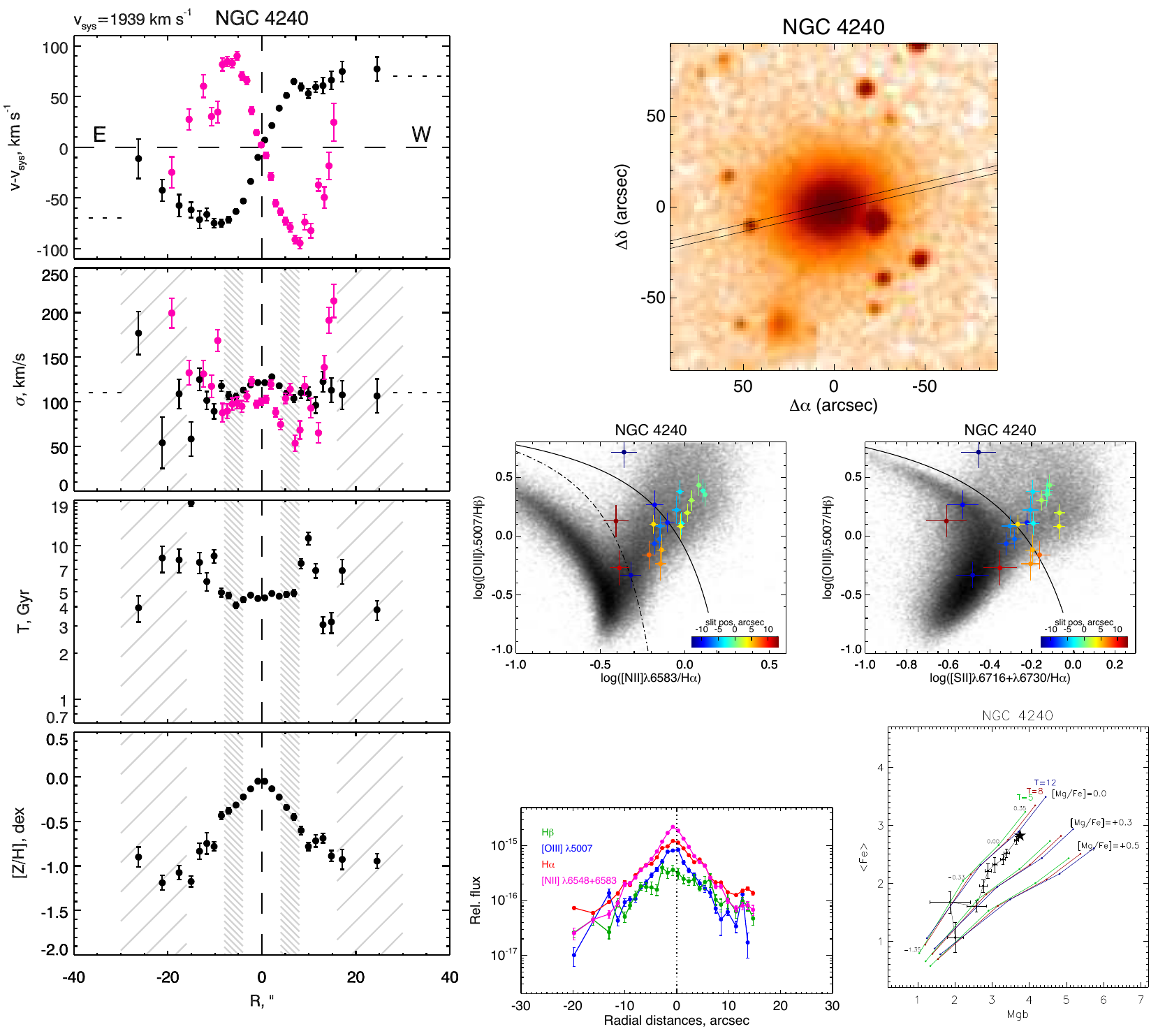}
}
\caption{NGC~4240. The same as previous figure.}
\end{figure} 

%%%%%%%%%%%%%%%%%%%%%%%%%%%%%%%%%%%%%%%%%%%%%%%%%%%%%%%%%%%%%%%%%%%%%%%%%%%%%%%%%%%%%%%%%%%%%%%%%%%%%%
\newpage
\begin{figure}
\centerline{
	\includegraphics[width=\textwidth]{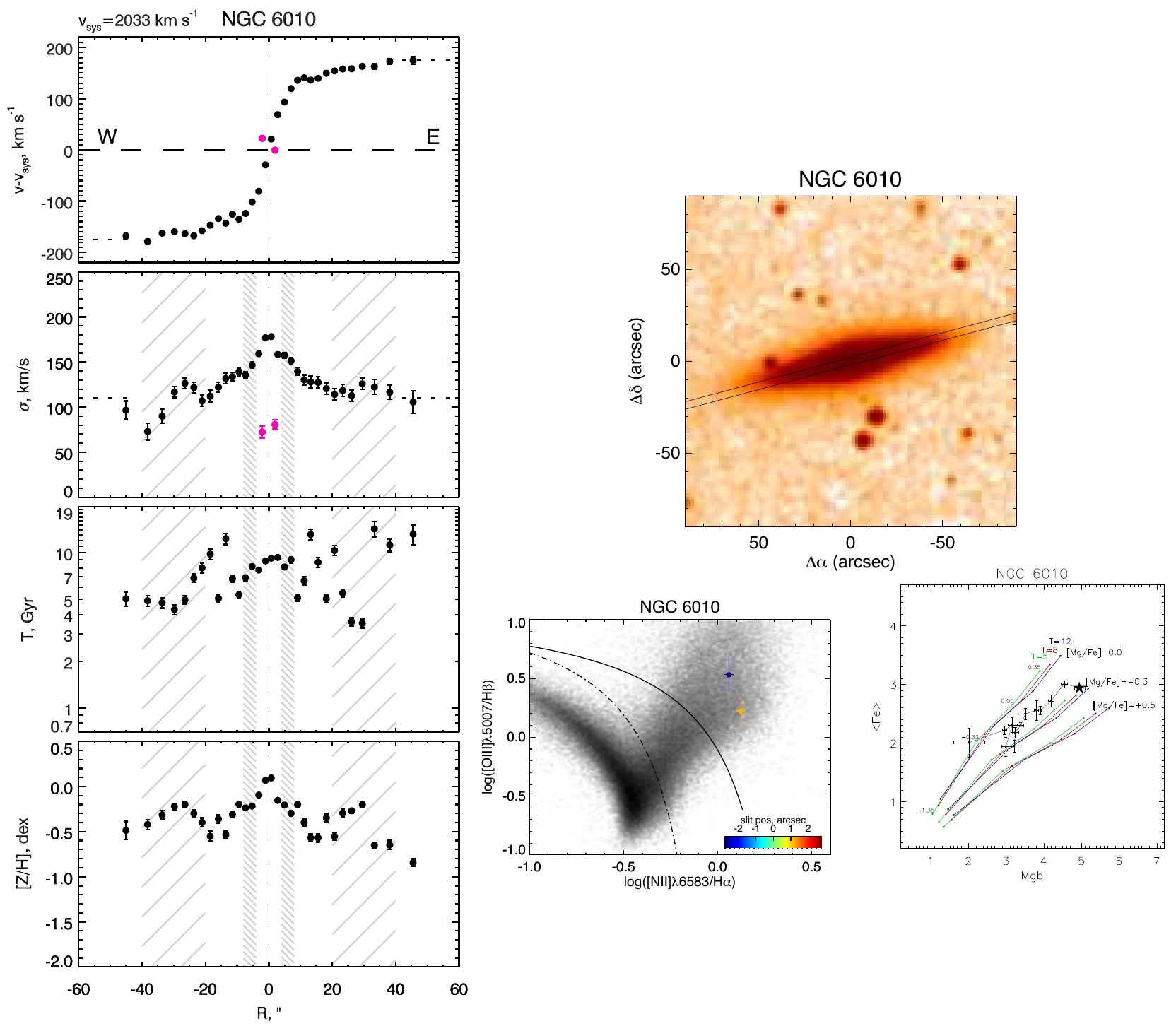}
}
\caption{NGC~6010. The same as previous figure.}
\label{fig_n6010}
\end{figure} 

%%%%%%%%%%%%%%%%%%%%%%%%%%%%%%%%%%%%%%%%%%%%%%%%%%%%%%%%%%%%%%%%%%%%%%%%%%%%%%%%%%%%%%%%%%%%%%%%%%%%%%
\newpage
\begin{figure}
\centerline{
	\includegraphics[width=\textwidth]{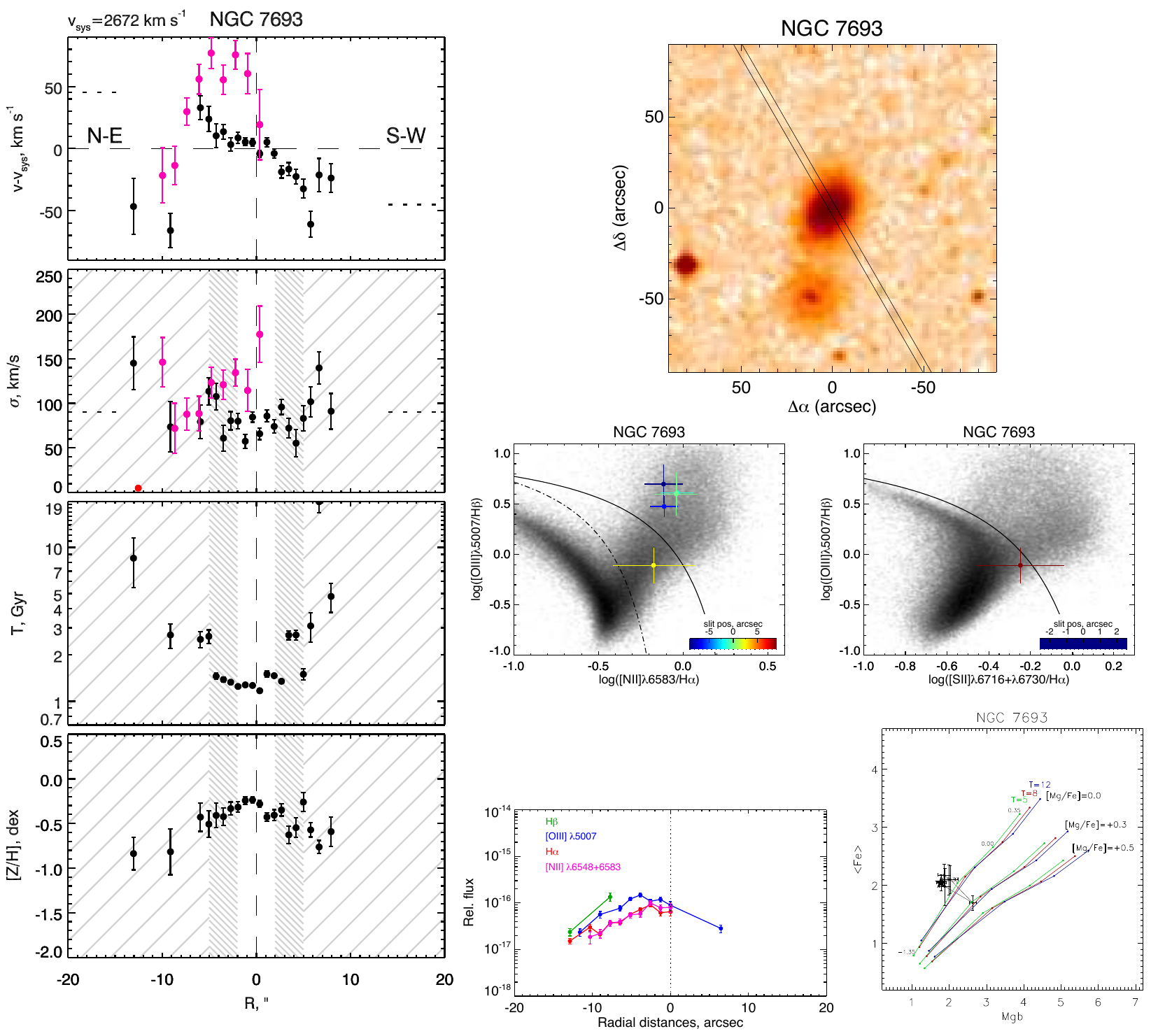}
}
\caption{NGC~7693. The same as previous figure.}
\end{figure} 
%%%%%%%%%%%%%%%%%%%%%%%%%%%%%%%%%%%%%%%%%%%%%%%%%%%%%%%%%%%%%%%%%%%%%%%%%%%%%%%%%%%%%%%%%%%%%%%%%%%%%%
\newpage
\begin{figure}
\centerline{
	\includegraphics[width=\textwidth]{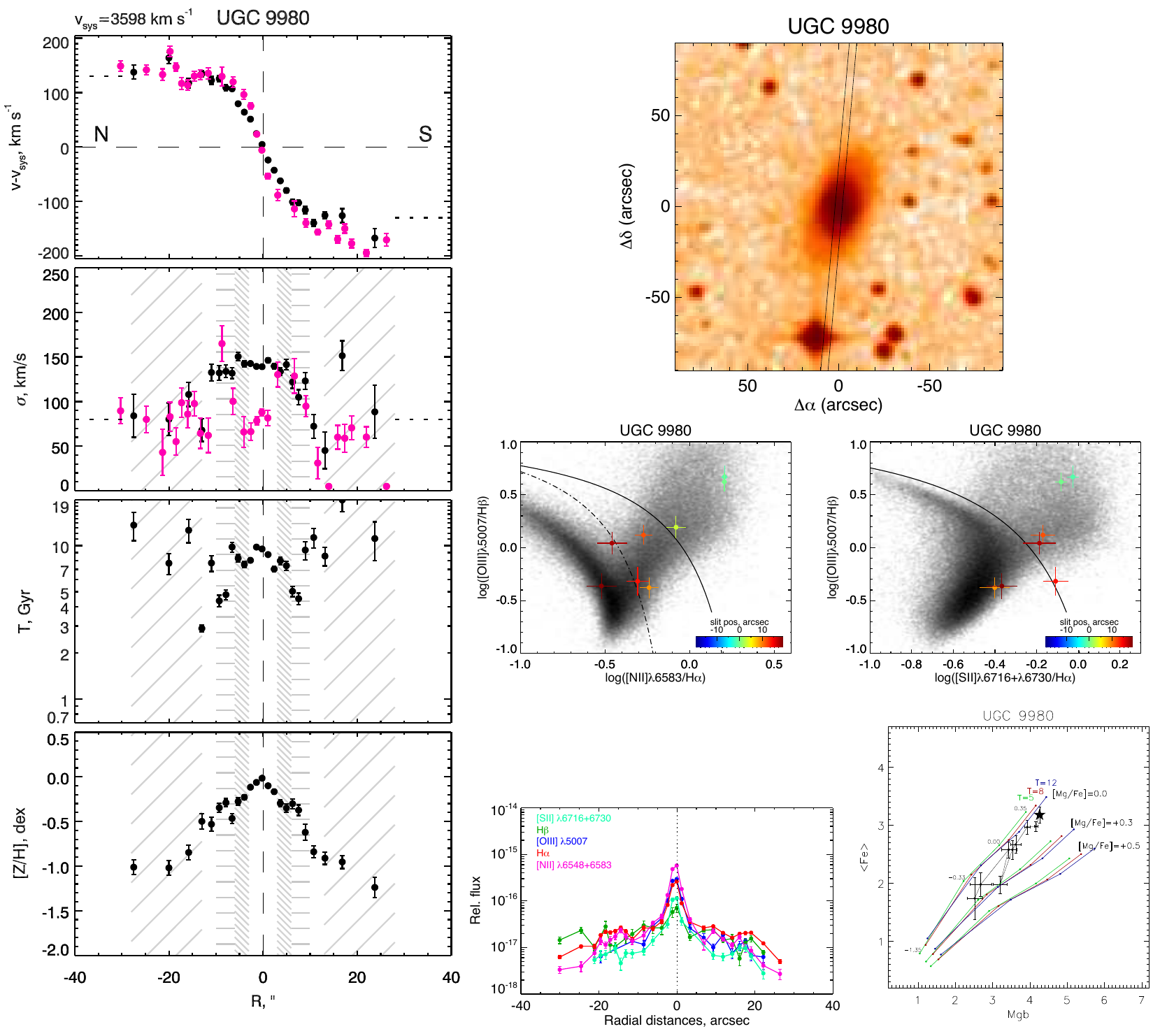}
}
\caption{UGC~9980. The same as previous figure.}
\end{figure} 
%%%%%%%%%%%%%%%%%%%%%%%%%%%%%%%%%%%%%%%%%%%%%%%%%%%%%%%%%%%%%%%%%%%%%%%%%%%%%%%%%%%%%%%%%%%%%%%%%%%%%%

\label{lastpage}

\end{appendix}

\end{document}